\documentclass[10pt]{article}

\usepackage{amssymb,amsmath,amsthm,mathrsfs} 

\usepackage[authoryear,sort&compress]{natbib}
\usepackage{colordvi,color,pspicture}
\usepackage{rotating}

\usepackage{setspace}

\newcommand{\X}{\mathcal{X}}
\newcommand{\U}{\mathcal{U}}
\newcommand{\I}{\mathbf{I}}

\newcommand{\R}{\mathbb{R}}

\newcommand{\E}{\mathbf{E}}
\newcommand{\Var}{\mathbf{Var}}
\newcommand{\Cov}{\mathbf{Cov}}

\newcommand{\ve}{\varepsilon}
\newcommand{\lam}{\lambda}
\newcommand{\al}{\alpha}

\newcommand{\ah}{\widehat{\alpha}}

\DeclareMathOperator{\BIN}{BIN}

\theoremstyle{plain}
\newtheorem{theorem}{Theorem}[section]

\theoremstyle{definition}

\theoremstyle{remark}

\newtheorem{remark}[theorem]{Remark}

\begin{document}

\title{Technical Report \# KU-EC-08-4:\\
On the Use of Nearest Neighbor Contingency Tables for Testing Spatial Segregation}
\author{
Elvan Ceyhan
\thanks{Department of Mathematics, Ko\c{c} University, 34450 Sar{\i}yer, Istanbul, Turkey}
}
\date{\today}
\maketitle

\begin{singlespace}
\begin{abstract}
For two or more classes (or types) of points, nearest neighbor contingency
tables (NNCTs) are constructed using nearest neighbor (NN) frequencies
and are used in testing spatial segregation of the classes.
Pielou's test of independence, Dixon's cell-specific, class-specific, and overall tests are
the tests based on NNCTs (i.e., they are NNCT-tests ).
These tests are designed and intended for use under the null pattern of random labeling (RL)
of completely mapped data.
However, it has been shown that Pielou's test is not appropriate for testing segregation
against the RL pattern while Dixon's tests are. 
In this article, we compare Pielou's and Dixon's NNCT-tests;
introduce the one-sided versions of Pielou's test;
extend the use of NNCT-tests for testing complete spatial randomness (CSR)
of points from two or more classes (which is called \emph{CSR independence}, henceforth).
We assess the finite sample performance of the tests
by an extensive Monte Carlo simulation study
and demonstrate that Dixon's tests are also appropriate for testing CSR independence;
but Pielou's test and the corresponding one-sided versions are
liberal for testing CSR independence or RL.
Furthermore, we show that Pielou's tests are only appropriate when
the NNCT is based on a random sample of (base, NN) pairs.
We also prove the consistency of the tests under their appropriate null hypotheses.
Moreover, we investigate the edge (or boundary) effects on the NNCT-tests
and compare the buffer zone and toroidal edge correction methods for these tests.
We illustrate the tests on a real life and an artificial data set.
\end{abstract}

\noindent
{\small {\it Keywords:} Association; completely mapped data;
complete spatial randomness; edge correction; random labeling; spatial point pattern

\vspace{.25 in}

$^*$corresponding author.\\
\indent {\it e-mail:} elceyhan@ku.edu.tr (E.~Ceyhan) }

\end{singlespace}

\newpage

\section{Introduction}
\label{sec:intro}
The analysis of spatial point patterns in
natural populations has been extensively studied in various fields.
In particular, spatial patterns in epidemiology, population biology, and ecology
have important implications.
A spatial point pattern includes the locations of some measurements,
such as the coordinates of trees in a region of interest.
These locations are referred to as \emph{events} by some authors,
in order to distinguish them from arbitrary points in the region of interest (\cite{diggle:2003}).
However in this article such a distinction is not necessary, 
as we only consider the locations of events.
Hence \emph{points} will refer to the locations of events, henceforth.
Most point patterns also include other types of measurements for each point,
such as categorical label (e.g., species label) or size (e.g., height of pine saplings).
Such labelled data are marked point patterns generated by marked point processes,
which define the distributions of the ``marks" or ``labels" to the locations
of the points and perhaps are the most common spatial point patterns.
For a general discussion of marked point processes, see \cite{diggle:2003},
\cite{gavrikov:1995}, \cite{penttinen:1992}, and \cite{schlather:2004}.
For convenience and generality, we call the different types of
points as ``classes".
From the early days on, the related research
has mostly been on only one class at a time;
i.e., on spatial pattern of each class (e.g., density, clumpiness, etc.).
These patterns in a one-class framework fall
under the pattern category called {\em spatial aggregation}
(\cite{coomes:1999}) or \emph{clustering}.
However, it is also of
practical interest to investigate the patterns of one class with
respect to the other classes (\cite{pielou:1961}).
The spatial relationships between two or more classes have important consequences
especially for plant species.
See, for example, \cite{pielou:1961},
\cite{dixon:1994}, and \cite{dixon:NNCTEco2002} for more detail.
Although we refer to types of points as ``classes", the ``class" can
be replaced by any characteristic of an observation at a particular location.
For example, the pattern of spatial segregation has been
investigated for species (\cite{diggle:2003}), age classes of plants
(\cite{hamill:1986}) and sexes of dioecious plants (\cite{nanami:1999}).
We also note that many of the epidemiological
applications are for a two-class system of case/control labels (\cite{waller:2004}).

In various fields, there are many tests available for spatial point patterns.
An extensive survey is provided by Kulldorff who
enumerates more than 100 such tests, most of which need adjustment
for some sort of inhomogeneity (\cite{kulldorff:2006}).
He also provides a general framework to classify these tests.
The most widely used tests include Pielou's test of segregation for two
classes (\cite{pielou:1961}) due to its ease of computation and
interpretation and Ripley's $K$ or $L$-functions (\cite{ripley:2004}).
The abundance of tests results because
(i) the tests for which Monte Carlo critical values are the only criteria
receive wide acceptance in various fields;
(ii) there are many different types of segregation patterns and some
tests are designed to detect only certain types of segregation patterns; and
(iii) the lack of cross-fertilization between different scientific
fields so that new tests are proposed unbeknownst to the developers of similar tests.

Nearest neighbor (NN) methods for spatial patterns include at least six
different groups (see, e.g., \cite{dixon:EncycEnv2002}).
The methods utilize some measure of (dis)similarity between a point and its NN;
such as the distance between the points or the class types of the points.
The latter type of similarity is used in the NN methods in this article.
Nearest neighbor contingency tables (NNCTs) are constructed using the NN
frequencies of classes and are used in testing spatial patterns.
\cite{pielou:1961} proposed tests (for segregation, symmetry, niche
specificity, etc.) based on NNCTs
under the RL of locations in the study region
and Dixon devised cell-specific, class-specific,
and overall tests based on NNCTs for the
two-class case (\cite{dixon:1994}) and extended his methodology to
the multi-class case (\cite{dixon:NNCTEco2002}) under RL.
Pielou's tests have been used for the two-class case only.
However it has been demonstrated that Pielou's test is not appropriate
for the NNCTs constructed under the RL of points (\cite{meagher:1980}).

In this article, we discuss the tests of spatial segregation based on NNCTs.
We describe the assumptions and hypotheses, the tests, and the
underlying sampling frameworks for Pielou's and Dixon's tests.
We propose one-sided versions of Pielou's test to detect the direction
of deviation from the RL pattern;
then extend the use of Pielou's and Dixon's tests for the CSR of points from two or more classes
in the region of interest.
However, we demonstrate that under CSR independence, Dixon's tests are conditional tests,
and propose a method to remove this conditional nature of Dixon's test.
We also compare the empirical sizes of the NNCT-tests with an extensive Monte Carlo simulation study,
where we demonstrate that Pielou's test and the corresponding one-sided versions are liberal
for rejecting RL or CSR independence, while Dixon's tests are about the desired nominal level.
We also prove the consistency of the tests under their appropriate null hypotheses;
show that Pielou's test is only appropriate when the NNCT
is based on a random sample of (base, NN) pairs (which is not realistic in practical situations).
We also investigate the edge (or boundary) effects on the NNCT-tests under CSR independence only
(since edge effects is not a concern under RL).

We describe the spatial point patterns of RL and
CSR independence in Section \ref{sec:spatial-pattern};
describe the NNCT-tests in Section \ref{sec:NNCT-tests}, in particular
we describe the construction of the NNCTs in Section \ref{sec:NNCT},
Pielou's test in Section \ref{sec:pielou's-test},
Dixon's NNCT-tests in Section \ref{sec:dixon-NNCT},
extend Dixon's test for the CSR independence pattern in Section \ref{sec:dixon-under-CSR}.
We prove the consistency of the NNCT-tests in Section \ref{sec:consistency}
(and defer the proofs to the Appendix Section);
present our extensive Monte Carlo simulation analysis in Section \ref{sec:Monte-Carlo},
in particular we compare the empirical significance levels of the tests under RL
in Section \ref{sec:Monte-Carlo-RL},
under CSR independence in Section \ref{sec:Monte-Carlo-CSR},
under the independence of rows and cell counts in NNCTs in Section \ref{sec:Monte-Carlo-Independence}.
We also consider the edge correction methods under the CSR independence pattern
in Section \ref{sec:edge-correct};
illustrate our methods on two example data sets in Section \ref{sec:examples}.
We provide our discussions and conclusions as well as guidelines for using the tests
in Section \ref{sec:disc-conc}.

\section{Spatial Point Patterns}
\label{sec:spatial-pattern}
For simplicity, we describe the spatial point patterns for two-class populations;
the extension to the multi-class case is straightforward.

In the univariate (i.e., one-class) spatial point pattern analysis,
the null hypothesis is usually \emph{complete spatial randomness} (\emph{CSR}) (\cite{diggle:2003}).
Given a spatial point pattern $\mathcal P=\{X_i\cdot \I(X_i \in D), i=1,\ldots,n: D \subset \mathbb{R}^2 \}$
where $X_i$ stands for the location of event $i$ (i.e., point $i$)
and $\I(X_i \in D)$ is the indicator function
which denotes the Bernoulli random variable denoting the event that
point $i$ is in region $D$.
The pattern $\mathcal P$ exhibits CSR if
given $n$ events (i.e., locations of the points) in domain $D$,
the events are an independent random sample from the uniform distribution on $D$.
Note that this condition also implies that
there is no spatial interaction;
i.e., the locations of these points have no influence on one another.
Furthermore, when the reference region $D$ is large,
the number of points in any planar region with area
$A(D)$ follows (approximately) a Poisson distribution with
intensity (i.e., number of points per unit area) denoted by $\lambda$
and mean $\lambda \cdot A(D)$.

To investigate the spatial interaction between two or more classes
in a bivariate process, usually there are two benchmark hypotheses:
(i) independence, which implies two classes of points are generated by a pair of independent
univariate processes and
(ii) random labeling (RL), which implies that the class labels
are randomly assigned to a given set of locations in the region of interest (\cite{diggle:2003}).
In this article, we will consider two random pattern types as our null hypotheses:
CSR of points from two classes
(this pattern will be called CSR independence, henceforth) or RL.

In CSR independence, points from each of the two classes satisfy the CSR pattern
in the region of interest.
On the other hand, RL is the pattern in which,
given a fixed set of locations in a region,
class labels are assigned to these fixed
locations randomly so that the labels are independent of the locations.
So, RL is less restrictive than CSR independence,
in the sense that RL does not impose any restrictions
on the distribution of the locations of the events,
but CSR independence is a process defining the spatial distribution of event locations.
The RL or CSR independence patterns imply a more refined null hypothesis
for the NNCT-tests,
namely, $H_o: \text{randomness in the NN structure}$.
%
When the points from each
class are assumed to be uniformly distributed over the region of interest,
then randomness in the NN structure is implied by the CSR independence pattern,
which is also referred to as (a type of) ``population independence" by some authors (\cite{goraud:2003}).
Note that this type of CSR is equivalent
to the case where the RL procedure is applied to a
given set of points from a CSR pattern
in the sense that after points are generated uniformly in the region,
the class labels are assigned randomly.
When only the labeling of a set of fixed points
(the allocation of the points could be regular, aggregated, or clustered, or of lattice type)
is random, the randomness in the NN structure is implied by RL pattern.

The distinction between the RL and CSR independence is very important
when defining the appropriate null model which depends on the particular ecological context.
\cite{goraud:2003} discuss the differences between independence and RL patterns
and show that the incorrect specification of the null pattern may result in
incorrect results, e.g., for Ripley's $K$ or $L$-functions.
They also propose some guidelines to determine
which null hypothesis is appropriate for a given situation.
For the null case of CSR independence (just independence in \cite{goraud:2003})
the locations of the points from two classes are \emph{a priori}
the result of (perhaps) different processes (e.g., individuals of different species or age cohorts),
whereas for the null case of RL some processes affect \emph{a posteriori} the individuals of a single population
(e.g., diseased versus non-diseased individuals of a single species).
Notice that although CSR independence and RL are not same,
they lead to the same null model (i.e., randomness in NN structure) for tests using NNCT,
which does not require spatially-explicit information.

Deviations from the null patterns (RL or CSR independence) were
first called (positive or negative) \emph{segregation}.
In Pielou's approach, two classes can be described
as ``unsegregated" if the NN of an individual is as
likely to be of the same class as the other class; that is,
neither class has a tendency to occur in one-class clumps or clusters.
{\em Negative segregation} occurs if the NN of a point is
more likely to be from a different class than the class of the point.
{\em Positive segregation} occurs if the NN
of a point is more likely to be of the same class as the class of the point;
 i.e., the members of the same class tend to be clumped or clustered
 (see, e.g., \cite{pielou:1961}).
The concept of ``negative segregation" as
described above, is more commonly referred to as {\em association},
whereas ``positive segregation" is merely called {\em segregation}.
See, for example, \cite{cressie:1993} and \cite{coomes:1999} for more detail.
Two classes may exhibit many different forms of segregation (\cite{pielou:1961}).
Although it is not possible to list all segregation types,
existence of segregation can be tested by using NNCTs.
In the statistical literature, association in contingency tables
generally refers to \emph{categorical association}.
To avoid confusion between this general association and the spatial pattern of
association, we call the former as ``categorical association" and
the latter as ``spatial association".
No such confusion occurs for segregation.

\section{Tests Based on Nearest Neighbor Contingency Tables}
\label{sec:NNCT-tests}
In this section, we present the construction of NNCTs and
then Pielou's and Dixon's tests based on NNCTs.

\subsection{Construction of Nearest Neighbor Contingency Tables}
\label{sec:NNCT}
Consider two classes labeled as $\{1,2\}$. NNCTs
are constructed using NN frequencies for each class.
Let $n_i$ be the number of points from class $i$ for $i \in \{1,2\}$ and $n=n_1+n_2$.
If we record the class of each point and
its NN, the NN relationships fall into
4 categories:
$(1,1),\,(1,2);\,(2,1),\,(2,2),$
where in cell $(i,j)$, class $i$ is the base class, while class $j$
is the class of its NN.
That is, a (base, NN) pair is
categorized according to its label.
Denoting $N_{ij}$ as the
observed frequency of cell $(i,j)$ for $i,j \in \{1,2\}$,
we obtain the NNCT in Table \ref{tab:NNCT-2x2} where $C_j$ is the sum of
column $j$; i.e., number of times class $j$ points serve as nearest
neighbors for $j\in\{1,2\}$.
Note also that $n=\sum_{i,j}N_{ij}$,
$n_i=\sum_{j=1}^2\, N_{ij}$, and $C_j=\sum_{i=1}^2\, N_{ij}$.
We adopt the convention that capital letters stand for random quantities,
while lower case letters stand for fixed quantities.

Let $\pi_{ij}=P(U=i,V=j)$ be the probability that the
pair of points $(U,V)$ falls in cell $(i,j)$;
i.e., the point $V$ is from class $j$ and is a NN of the point $U$ which is from class $i$.
Furthermore, let $\nu_i=\pi_{i1}+\pi_{i2}$ for $i=1,2$, that is, the
probability of a base point to be of class $i$.
Similarly, let
$\kappa_j=\pi_{1j}+\pi_{2j}$ for $j=1,2$, that is, the probability
of a NN point to be of class $j$.
The sample versions of these probabilities are $\widehat{\pi}_{ij}=N_{ij}/n$
for $\pi_{ij}$, $\widehat{\nu}_i=n_i/n$ for $\nu_i$, and
$\widehat{\kappa}_j=C_j/n$ for $\kappa_j$.

A (base, NN) pair can be categorized as reflexive or non-reflexive,
regardless of the classes of the members of the pair.
For a (base, NN) pair, $(X,Y)$,
(i.e., $Y$ is a NN of $X$), if $X$ is a NN of
$Y$ (i.e., $(Y,X)$ is also a (base, NN) pair), then the pair $(X,Y)$
is called {\em reflexive}.
If a (base, NN) pair is not reflexive,
then it is a \emph{non-reflexive} pair.
Moreover, a point can
serve as NN to none or several other points. That is, a point can be a
shared NN to $k~(k < 6)$ other points in $\R^2$
(\cite{clark:1955}).

\subsection{Pielou's Test of Segregation}
\label{sec:pielou's-test}
Pielou constructed NNCTs based on NN frequencies
which yield tests that are independent of quadrat size
(\cite{pielou:1961}, \cite{krebs:1972}).
In the two-class case, Pielou used the usual Pearson's $\chi^2$ test
of independence (with 1 df) to test for presence or lack of
segregation (\cite{pielou:1961}).
Due to the ease in computation and interpretation, her test of
segregation is widely used in ecology  (\cite{meagher:1980}) for
both completely mapped or sparsely sampled data.
In particular, Pielou has described and used her test of segregation
for \emph{completely mapped data}, although her test is
not appropriate for such data (see \cite{meagher:1980} and \cite{dixon:1994}).
A data set is {\em completely mapped}, if the points (i.e., locations of all events)
in a defined space are observed.
Alternatively, \emph{sparse sampling} might be
suitable for the use of Pielou's test as suggested by \cite{dixon:1994}.
Although sparse sampling is not clearly defined in the literature,
it can be classified into two types.
The sparse sampling schemes depend on the events (members of a class occupying
a location) or arbitrary points in the region of interest.
The most well known sparse sampling method is the \emph{quadrat sampling},
in which the number of events falling into each of several (preferably random)
small subregions (quadrats) is recorded.
However construction of NNCTs may not even be possible in such a scheme,
since the NN information may be lost
when only the number of events are recorded for each quadrat.
The second sparse sampling scheme is the \emph{distance sampling},
in which the basic unit is an arbitrary point
(not necessarily from the events) and the information based on the
distance to the nearest event is recorded (\cite{solow:1989}).
Notice that this type of distance sampling scheme does not yield
sufficient information for the construction of NNCTs.

Pearson's $\chi^2$ test of independence for a $2\times 2$ contingency table,
in general, can be assumed to develop from one of the following frameworks:
\emph{Poisson, row-wise binomial}, or \emph{overall multinomial sampling frameworks}.
Below we briefly describe these frameworks for $2 \times 2$ contingency tables.
Let $\widetilde \pi_{ij}$ be the probability of a point to fall in cell $(i,j)$
in the contingency table,
and $\widetilde \nu_i$ and $\widetilde \kappa_j$ be
the probabilities that the point is of row category $i$ and of column
category $j$, respectively.

The test statistic for Pielou's test (which is same as Pearson's test) is given by
\begin{equation}
\label{eqn:piel-chisq-2x2}
\X^2_P=\sum_{i=1}^2\sum_{j=1}^2 \frac{(N_{ij}-\E[N_{ij}])^2}{\E[N_{ij}]}.
\end{equation}

\noindent \textbf{Poisson Sampling Framework:}
Each category count in the contingency table is
assumed to be an independent Poisson variate.
Another key feature is the independence of cell counts,
which would imply that $\widetilde \pi_{ij}=\widetilde \nu_i\,\widetilde \kappa_j$
for all $(i,j) \in \{1,2\}$.
This independence can be tested by Pearson's $\chi^2$ test
for large samples and Fisher's exact test or the
exact version of the Pearson's test for small samples (\cite{agresti:1992}).
The null hypothesis in this framework is
$H_o:\widetilde \pi_{ij}=\widetilde \nu_i\,\widetilde \kappa_j$,
and $\E[N_{ij}]$ of Equation \eqref{eqn:piel-chisq-2x2} is $\frac{N_i\,C_j}{n}$.
Among the alternatives,
$H_a:\widetilde \pi_{ii}>\widetilde \nu_i\,\widetilde \kappa_i$
is suggestive of positive categorical association for classes $i=1,2$,
while $H_a:\widetilde \pi_{12}<\widetilde \nu_1\,\widetilde \kappa_2$ or
$H_a:\widetilde \pi_{21}<\widetilde \nu_2\,\widetilde \kappa_1$
is suggestive of negative categorical association between classes 1 and 2.

However, in the case of NNCTs, cell counts,
e.g., $N_{12}$ and $N_{21}$ are not independent under RL or CSR independence.
Because, a (base, NN) pair is more likely to be a reflexive pair,
rather than a non-reflexive pair under RL or CSR independence (\cite{meagher:1980}).
Thus under Poisson sampling framework,
Pielou's test would be inappropriate for testing RL or CSR independence.

\noindent \textbf{Row-wise Binomial Sampling Framework:}
In this framework, we assume that $N_i=n_i$ are given and $N_{ij} \sim
\BIN(n_i,\widetilde \pi_{ij})$, the binomial distribution with $n_i$ trials and
probability of success being $\widetilde \pi_{ij}$. Notice that for more than two
classes, this will be \emph{row-wise multinomial framework}.

Then the null hypothesis for this test is $H_o: \widetilde \pi_{11}=\widetilde \pi_{21}$
which also implies $\widetilde \pi_{12}=\widetilde \pi_{22}$.
The alternative $H_a: \widetilde \pi_{11}>\widetilde \pi_{21}$
would correspond to positive categorical association
which would also imply $\widetilde \pi_{22}>\widetilde \pi_{12}$.
Similarly, the alternative
$H_a: \widetilde \pi_{11}<\widetilde \pi_{21}$ would correspond
to negative categorical association which would also imply
$\widetilde \pi_{22}<\widetilde \pi_{12}$.

Under $H_o$, we can parametrize the null model as $N_{ij}\sim \BIN(n_i,\widetilde \kappa_j)$
where $\widetilde \kappa_j$ can be estimated as $C_j/n$ and is assumed to equal
the expectation $\E[C_j/n]$.
Then $H_o: \widetilde \pi_{11}=\widetilde \pi_{21}=\widetilde \kappa_1$ is equivalent to
$H_o: \E[N_{11}/n_1]=\E[N_{21}/n_2]=\widetilde \kappa_1$ which is equivalent to
$H_o: \E[N_{11}]=n_1\,\widetilde \kappa_1 \text{ and } \E[N_{21}]=n_2\,\widetilde \kappa_1$
which, for large $n,\,n_1,$ and $n_2$, is equivalent to
$H_o: \E[N_{11}/n]=n_1\,\widetilde \kappa_1/n=\widetilde \nu_1\,\widetilde \kappa_1
\text{ and }
\E[N_{21}/n]=n_2\,\widetilde \kappa_1/n=\widetilde \nu_2\,\widetilde \kappa_2$.

Under $H_o$, if $\widetilde \kappa_j$ are known, $\X^2_P$ is approximately distributed
as $\chi^2_2$ (i.e., $\chi^2$ distribution with 2 degrees of freedom) for large $n_i$;
if $\widetilde \kappa_j$ are not known, but estimated as $C_j/n$,
then $\X^2_P$ is approximately distributed as $\chi^2_1$ for large $n_i$.
In most practical situations, the latter case will occur,
so $\chi^2_1$ distribution is used for this test.

In the two-class case, $(N_{11},N_{12})$ and $(N_{21},N_{22})$
are assumed to be independent and so are the individual trials,
namely, (base, NN) pairs.
Under RL or CSR independence, this assumption is invalid for completely mapped data.
Because the trials that constitute $N_{ii}$ for $i=1,2$
are not independent due to reflexivity and shared NN structure;
likewise, $N_{12}$ and $N_{21}$ are not independent.
Hence, Pielou's test is not appropriate for RL or CSR independence in this framework either.

\noindent \textbf{Overall Multinomial Sampling Framework:}
An alternative sampling framework for contingency tables,
in general, is that the cell counts are assumed to be
from independent multinomial trials.
That is, for the two-class case,
$$\mathbf{N}=(N_{11},N_{12},N_{21},N_{22}) \sim
\mathscr M(n,\widetilde \pi_{11},\widetilde \pi_{12},\widetilde \pi_{21},\widetilde \pi_{22})$$ hence the name {\em
overall multinomial framework}.
The null hypothesis in this framework is $H_o:(\widetilde \pi_{11},\widetilde \pi_{12})=(\widetilde \pi_{21},\widetilde \pi_{22})$
and $\E[N_{ij}]$ in Equation \eqref{eqn:piel-chisq-2x2} is $N_i\,C_j/n$.
The multinomial counts are not independent (since they are negatively correlated)
when conditioned on their total.
This dependence alleviates as the sample size increases,
but might confound the small sample results.
In addition to this mild dependence,
the NNCT cell counts are not independent due to, e.g., reflexivity.
Hence the overall multinomial framework is not appropriate for
NNCTs based on RL or CSR independence either.

Note that conditional on $N_i=n_i$, the overall multinomial
framework reduces to the row-wise multinomial framework.
Furthermore, when the parameters are not known but estimated from
the marginal sums, all frameworks yield tests that are approximately
distributed as $\chi^2_1$ for large $n$.

\subsubsection{One-Sided Versions of Pielou's Test of Segregation}
\label{sec:one-sided-pielou-tests}
Pielou's test is a general two-sided test,
hence it does not indicate the direction of the deviation
(e.g., positive or negative categorical association) from the null case.
To determine the direction, one needs to check the NNCT.
Since $\X_P^2 \stackrel{approx}{\sim}
\chi^2_1$, for large $n$, we can write $\X_P^2=Z_n^2$ where $Z_n
\stackrel{approx}{\sim} N(0,1)$, the standard normal distribution.
By some algebraic manipulations, among other possibilities, for the
row-wise multinomial framework, $Z_n$ can be written as
\begin{equation}
\label{eqn:2x2-Zform}
Z_n=\left(\frac{N_{11}}{n_1}-\frac{N_{21}}{n_2} \right)\sqrt{\frac{n_1\,n_2\,n}{C_1\,C_2}}.
\end{equation}
See (\cite{bickel:1977}) for the sketch of the derivation.
Positive values of $Z_n$ indicate positive categorical association,
while negative values indicate negative categorical association.
When cell counts are independent,
a reasonable $\al$-level test is rejecting $H_o$ if
$Z_n>z_{1-\al}$ for segregation or if $Z_n<z_{\al}$ for spatial association.
The $\al$-level $\chi^2$ test in which we
reject for $\X_P^2>\chi^2_1(1-\al)$ is equivalent to the two-sided
$\al$-level test based on $Z_n$.

The corresponding test statistic for the overall multinomial framework can be written as
\begin{equation}
\label{eqn:Zn-tilde}
\widetilde{Z}_n=\left( N_{11}-\frac{n_1\,c_1}{n} \right) \sqrt{\frac{n^3}{n_1\,n_2\,c_1\,c_2}}.
\end{equation}

Once again, we point out that these one-sided tests are not
appropriate for testing RL or CSR independence,
due to inherent dependence of cell counts in NNCTs based on such patterns.

\begin{remark}
\label{rem:pielou-appropriate-null}
\textbf{Appropriate Null Case for Pielou's Tests:}
In Pielou's test, each of the Poisson and row-wise
binomial sampling frameworks for cell counts
assumes that the trials (i.e., the cross-categorization of base-NN pairs) are independent
and in the overall multinomial framework,
there is mild dependence between the cell counts.
The independence of rows and individual trials (i.e., cells) would follow
if NNCT were based on a random sample of (base label, NN label).
But unfortunately, this usually is not realistic in practice,
although it might have theoretical appeal.
When we have a random sample of (base label, NN label) pairs
(which are also called the (base, NN) pairs),
the null hypothesis for Pielou's test is equivalent to the case that
the vector of probabilities for the cell frequencies for each row are identical.
Hence if the NNCT is based on a random sample of (base, NN) pairs,
then any of the sampling frameworks would be appropriate which in turn implies
the appropriateness of Pearson's test of independence for the NNCT.
The null hypothesis in each of the sampling frameworks will
imply independence between the patterns of the two classes.
On the other hand the alternatives of positive categorical association will correspond to
segregation of the classes,
while negative categorical association will correspond to
spatial association of the classes.
$\square$
\end{remark}

\subsection{Dixon's NNCT-Tests}
\label{sec:dixon-NNCT}
Dixon proposed a series of tests for segregation based on NNCTs, namely,
cell- and class-specific tests, and overall test of segregation
under RL (\cite{dixon:1994}).

\subsubsection{Dixon's Cell-Specific Tests}
\label{sec:dixon-cell-spec}
The level of segregation is estimated by
comparing the observed NNCT cell counts to the expected NNCT cell counts under
RL of fixed points. 
Dixon demonstrates that under RL, one can write down the cell
frequencies as Moran join-count statistics (\cite{moran:1948}).
He then derives the means, variances, and covariances of the cell
counts (i.e., frequencies) (\cite{dixon:1994} and \cite{dixon:NNCTEco2002}).

When the null hypothesis is RL, we have
\begin{equation}
\label{eqn:Exp[Nij]}
\E[N_{ij}]=
\begin{cases}
n_i(n_i-1)/(n-1) & \text{if $i=j$,}\\
n_i\,n_j/(n-1)   & \text{if $i\not=j$,}
\end{cases}
\end{equation}
or equivalently
$$\pi_{ij}=\frac{n_{i}\,(n_i-1)}{n(n-1)}\I(i=j)+\frac{n_{i}\,n_j}{n(n-1)}\I(i\not=j).$$

The test statistic suggested by Dixon is given by
\begin{equation}
\label{eqn:dixon-Zij}
Z^D_{ij}=\frac{N_{ij}-\E[N_{ij}]}{\sqrt{\Var[N_{ij}]}}
\end{equation}
where
{\small
\begin{equation}
\label{eqn:VarNij}
\Var[N_{ij}]=
\begin{cases}
(n+R)p_{ii}+(2n-2R+Q)p_{iii}+(n^2-3n-Q+R)p_{iiii}-(np_{ii})^2 & \text{if $i=j$,}\\
n\,p_{ij}+Q\,p_{iij}+(n^2-3\,n-Q+R)\,p_{iijj} -(n\,p_{ij})^2               & \text{if $i \not= j$,}
\end{cases}
\end{equation}
}
with $p_{xx}$, $p_{xxx}$, and $p_{xxxx}$
are the probabilities that a randomly picked pair,
triplet, or quartet of points, respectively, are the indicated classes and
are given by
\begin{align}
\label{eqn:probs}
p_{ii}&=\frac{n_i\,(n_i-1)}{n\,(n-1)},  & p_{ij}&=\frac{n_i\,n_j}{n\,(n-1)},\nonumber\\
p_{iii}&=\frac{n_i\,(n_i-1)\,(n_i-2)}{n\,(n-1)\,(n-2)}, &
p_{iij}&=\frac{n_i\,(n_i-1)\,n_j}{n\,(n-1)\,(n-2)},\\
 p_{iijj}&=\frac{n_i\,(n_i-1)\,n_j\,(n_j-1)}{n\,(n-1)\,(n-2)\,(n-3)},&
p_{iiii}&=\frac{n_i\,(n_i-1)\,(n_i-2)\,(n_i-3)}{n\,(n-1)\,(n-2)\,(n-3)}.\nonumber
\end{align}
Furthermore, $Q$ is the number of points with shared NNs,
which occurs when two or more points share a NN and
$R$ is twice the number of reflexive pairs.
Then $Q=2\,(Q_2+3\,Q_3+6\,Q_4+10\,Q_5+15\,Q_6)$
where $Q_k$ is the number of points that serve
as a NN to other points $k$ times.

One-sided and two-sided tests are possible for each cell $(i,j)$
using the asymptotic normal approximation of $Z^D_{ij}$ given in
Equation \eqref{eqn:dixon-Zij} (\cite{dixon:1994}).
In Dixon's framework, $N_{ij}$ are random quantities;
and the quantities in the expectations,
hypotheses, and variances are conditional on $N=n$ and $N_i=n_i$ for
$i \in \{1,2\}$. The column sums are irrelevant for Dixon's tests.

We describe the setting in a broader context.
Let $\nu_i$ be the
probability of an arbitrary point being from class $i$.
Then under RL, $\pi_{ij}=\nu_i \, \nu_j$ and the expression
$\frac{n_{i}\,(n_i-1)}{n(n-1)}\I(i=j)+\frac{n_{i}\,n_j}{n(n-1)}\I(i\not=j)$
can be viewed as an estimate of $\pi_{ij}$ and denoted as $\widehat{\pi}_{ij}$.
Furthermore, given large $N=n$, under the null hypothesis of RL
the expected values given in Equation \eqref{eqn:Exp[Nij]} implies
$$H_o:\pi_{ij}=\nu_i \, \nu_j$$
and the test statistic $(N_{ij}/n-\pi_{ij})\big/\sqrt{\Var[N_{ij}]}$ is
approximately equivalent to $Z^D_{ij}$ in Equation \eqref{eqn:dixon-Zij}.
In Dixon's framework, for large $n$ and $n_i$,
the row marginals satisfy $\E[N_i/n]=\nu_i$ and the column marginals
satisfy $\E[C_j/n]=\kappa_j=\sum_{i=1}^2 \nu_i\,\nu_j=\nu_j$.

\subsubsection{Dixon's Overall Test of Segregation}
\label{sec:dixon-overall}
Dixon's overall test of segregation tests the hypothesis that
expected values of the cell counts in the NNCT are equal to
the ones given in Equation \eqref{eqn:Exp[Nij]}.
In the two-class case, he calculates
$Z_{ii}=(N_{ii}-\E[N_{ii}])/\sqrt{\Var[N_{ii}]}$ for both $i \in
\{1,2\}$ and then combines these test statistics into a statistic
that is asymptotically distributed as $\chi^2_2$ (\cite{dixon:1994}).
The suggested test statistic is given by
\begin{equation}
\label{eqn:dix-chisq-2x2}
\X_D^2=\mathbf{Y}'\Sigma^{-1}\mathbf{Y}=
\left[
\begin{array}{c}
N_{11}-\E[N_{11}] \\
N_{22}-\E[N_{22}]
\end{array}
\right]'
\left[
\begin{array}{cc}
\Var[N_{11}] & \Cov[N_{11},N_{22}] \\
\Cov[N_{11},N_{22}] & \Var[N_{22}] \\
\end{array}
\right]^{-1}
\left[
\begin{array}{c}
N_{11}-\E[N_{11}] \\
N_{22}-\E[N_{22}]
\end{array}
\right]
\end{equation}
where $\E[N_{ii}]$ are as in Equation \eqref{eqn:Exp[Nij]},
$\Var[N_{ii}]$ are as in Equation \eqref{eqn:VarNij},
and
$$\Cov[N_{11},\,N_{22}]=\left( n^2-3\,n-Q+R \right)\,p_{1122}-n^2\,p_{11}\,p_{22}.$$
Under $H_o:\E[N_{ii}/n]=\nu_i^2 \text{ for $i=1,2$}$,
and $\E[\X_D^2]=2$ and $\Var[\X_D^2]=4$; i.e., the non-centrality parameter $\lam=0$.
If we parametrize the segregation alternative as
$H^S_a:\,\E[N_{11}/n]=(\nu_1+\ve_1)^2 \text{ and } \E[N_{22}/n]=(1-\nu_1+\ve_2)^2$
for some $\ve_1,\ve_2 > 0$.
Then under $H^S_a$, the non-centrality parameter satisfies
$\lam=\lam(\ve_1,\ve_2)>0$ since $\lam(\ve_1,\ve_2)=
\E_S[\mathbf{Y}]'\Sigma_S^{-1}\E_S[\mathbf{Y}]$
where
$$
\frac{1}{n}\E_S[\mathbf{Y}]'=[(\nu_1+\ve_1),(1-\nu_1+\ve_2)]
$$
and $\Sigma_S$ is the (positive definite) variance-covariance matrix
of the cell counts under $H^S_a$.
If the association alternative is parametrized as above with $\ve_1,\ve_2 < 0$
then we obtain the same non-centrality parameter $\lam(\ve_1,\ve_2)$.

\cite{dixon:NNCTEco2002} extends his test for multi-class case
(i.e., for the case with three or more classes).
He also partitions the overall test statistic $\X_D^2$ into class-specific
test statistics each of which are dependent but
approximately follow a $\chi^2$ distribution.

\subsection{Dixon's Tests under CSR Independence}
\label{sec:dixon-under-CSR}
The expected values of the NNCT cell counts given in Equation \eqref{eqn:Exp[Nij]}
are derived under RL or CSR independence by \cite{dixon:1994}.
However, the variances and covariances of the cell
counts used in Sections \ref{sec:dixon-cell-spec}
and \ref{sec:dixon-overall} are derived under RL pattern only
(\cite{dixon:1994} and \cite{dixon:NNCTEco2002}).

When the null hypothesis is CSR independence,
the expressions for the variances and covariances
of the NNCT cell counts are as in RL case,
except they are conditional on $Q$ and $R$.
The quantities $Q$ and $R$ are fixed under RL,
but random under CSR independence.
Hence under CSR independence Dixon's cell-specific test given in Equation \eqref{eqn:dixon-Zij}
asymptotically has $N(0,1)$ distribution
and overall test given in Equation \eqref{eqn:dix-chisq-2x2}
asymptotically has $\chi^2_2$, conditional on $Q$ and $R$.
Under the CSR independence pattern, the unconditional variances and
covariances (hence the unconditional asymptotic distributions)
can be obtained by replacing $Q$ and $R$ with their expectations.

Unfortunately, given the difficulty of calculating the
expectations of $Q$ and $R$ under CSR independence,
it seems reasonable and convenient to use test statistics employing the
unconditional variances and covariances even when assessing their
behavior under the CSR independence pattern.
Alternatively, one can estimate the values of $Q$ and $R$ empirically,
and substitute these estimates in the variance and covariance expressions.
For example, for homogeneous planar Poisson pattern,
we have $\E[Q/n] \approx .632786$ and $\E[R/n] \approx 0.621120$
(estimated empirically by 1000000 Monte Carlo simulations
for various values of $n=n_1+n_2$ on the unit square).

To assess the influence of conditioning on
the performance of Dixon's tests for the two-class case,
we consider both the conditional version of these tests,
as well as the unconditional version, in which the terms
$Q$ and $R$ are replaced by $0.63 \, n$ and $0.62 \,n$, respectively.
We call the latter type of correction as \emph{QR-adjustment}
and the transformed tests as \emph{QR-adjusted} tests, henceforth.
QR-adjusted version of Dixon's cell-specific test statistic
for cell $(i,j)$ is denoted by $Z_{ij}^{D,qr}$
and of the overall test statistic is denoted by $\X^2_{D,qr}$.

\begin{remark}
\label{rem:extension-to-multi-class}
\textbf{Extension of NNCT-Tests to Multi-Class Case:}
Dixon has extended his tests
into multi-class situation with three or more classes
(\cite{dixon:NNCTEco2002}).
For $q$ classes with $q>2$, the corresponding NNCT is of dimension $q \times q$.
It is possible to define $q^2$ cell-specific tests as in  Equation \eqref{eqn:dixon-Zij},
and one can combine the tests into one overall test similar to the one
given in  Equation \eqref{eqn:dix-chisq-2x2}, which will have $\chi^2_{q(q-1)}$, asymptotically.
On the other hand, Pielou's test is
defined and has only been used for the two-class spatial patterns.
Its inappropriateness discourages its immediate extension to multi-class patterns.
$\square$
\end{remark}

\subsection{Comparison of Pielou's and Dixon's Tests}
\label{sec:comparison-pielou-vs-dixon}
Dixon points out two problems with Pielou's test of independence:
(i) it fails to identify certain types of segregation
(e.g., mother-daughter processes)
and
(ii) the sampling distribution of NNCT cell counts
is not appropriate (see \cite{dixon:1994}).
In a mother-daughter process,
mothers are distributed randomly in the region of interest,
while the daughters are randomly displaced within close vicinity
of their mothers.
In such a process, it is possible to obtain a NNCT in which
the cell counts are very similar to the ones expected under
the Pielou null hypothesis,
while in reality the pattern exhibits the segregation of the daughters.
For more detail on mother-daughter processes and examples
for which Pielou's test giving misleading results, see (\cite{dixon:1994}).
Problem (ii) was first noted by \cite{meagher:1980} who identify the main source
of it to be reflexivity of (base, NN) pairs.
As an alternative, they
suggest using Monte Carlo simulations for Pielou's test.
Dixon shows that Pielou's test is not appropriate for completely mapped data,
but suggests that it might be appropriate for sparsely sampled data (\cite{dixon:1994}).

In Pielou's test, each of the sampling frameworks requires
that the cell counts are independent.
However, when a trial is
label categorization of a (base, NN) pair, the assumption of
independence between trials is violated due to reflexivity and
shared NN structure.
Thus Pielou's test measures deviations not only
from the null pattern of RL or CSR independence but also from
the independence of trials.
This also suggests that Pielou's test would be liberal
in rejecting the null hypothesis.
The reflexivity and shared NN
structure are not merely finite sample patterns, as they follow a
certain non-degenerate distribution even when
$n \rightarrow \infty$ (\cite{clark:1955}).

By construction, Pielou's test is used to test independence of the class labels
of the (base, NN) pairs, but ignores the spatial information
(hence ignores the spatial dependence, e.g., reflexivity of NNs).
On the other hand, Dixon's tests are used for the null hypotheses
of RL or CSR independence and uses more of the spatial information.
For Dixon's tests, the underlying sampling framework for cell
counts is different from Poisson, row-wise binomial, or overall
multinomial sampling models of the contingency tables.
In his framework, the probability of class $j$ point serving as a
NN of a class $i$ point depends only on the class sizes
(i.e., row sums), but not the total number of times class $j$ serves as a
NN (i.e., column sums).
On the other hand, Pielou's test depends on both row and column sums.
In fact, Pielou starts her arguments with NN
probabilities depending on class sizes (row sums) in (\cite{pielou:1961}, pp 257-258).
Then she leaves this track of
development because of dependence due to shared NN structure (i.e.,
the distribution of $Q_k$ (\cite{clark:1955}).
For testing RL or CSR independence, Dixon's framework is more appropriate as a
sampling distribution for NNCT cell counts,
as it accounts for the inherent spatial dependence between observations.

\section{Consistency of the NNCT-Tests}
\label{sec:consistency}
The null hypotheses are different for Pielou's
and Dixon's framework of testing spatial patterns,
and so are the alternative hypotheses.
Hence the acceptance regions are different (see, e.g., \cite{dixon:1994}),
and no test is uniformly superior to the other,
since both $A_D \setminus A_P$ and $A_P \setminus A_D$ are non-empty,
where $A_D$ is the acceptance region for Dixon's test and
$A_P$ is the acceptance region for Pielou's test.
That is, there are situations in which Pielou's test yields a significant result,
while Dixon's test finds no significant segregation, and vice versa.
For example, a pattern resulting from a mother/daugter process can fall in $A_P \setminus A_D$
(see Section \ref{sec:comparison-pielou-vs-dixon}).
On the other hand, a process in which row and column sums in a NNCT are close
but the cell counts are different than expected under Pielou null hypothesis,
and similar to the ones expected under Dixon null hypothesis might yield
a pattern that falls in $A_D \setminus A_P$.
Therefore the comparison of the tests (even for large samples) is inappropriate.
But any reasonable test should have more power as the sample size increases.
So, we prove that the tests under consideration are
consistent, although they have appropriate size under different null hypotheses.
The proofs of lemmas and theorems in this section are
all deferred to the Appendix.

In the following theorems we use the consistency of tests
based on statistics that have $N(0,1)$ or $\chi_{\nu}^2$
distributions asymptotically.
First, we prove the consistency of Pielou's test of segregation
and the one-sided versions.
Let $z(\al)$ be the $100(1-\al)^{th}$ percentile for the standard normal distribution
and $\chi^2_{\nu}(\al)$ be the $100(1-\al)^{th}$ percentile for $\chi^2$ distribution
with $\nu$ df.

\begin{theorem}
\label{thm:piel-directional}

(I) Suppose the NNCT is constructed based on a random sample of
(base, NN) pairs. The test for segregation
$H^S_a:\widetilde \pi_{11}> \widetilde \pi_{22}$ (spatial association $H^A_a:\widetilde \pi_{11} <
\widetilde \pi_{22}$) which rejects
$H_o:\,\widetilde \pi_{11} = \widetilde \pi_{22}$ for 
$Z_n>z(1-\al)$ (for $Z_n<z(\al)$) with $Z_n$ given in Equation \eqref{eqn:2x2-Zform}
has size $\alpha$ and is consistent.
Likewise, the test against
segregation $H^S_a:\,\E[N_{11}/n]>\widetilde \nu_1\, \widetilde \kappa_1$ which rejects
$H_o:\, \E[N_{11}/n]=\widetilde \nu_1\, \widetilde \kappa_1$ for
$\widetilde{Z}_n>z(1-\al)$ with $\widetilde{Z}_n$ given in
Equation \eqref{eqn:Zn-tilde} has size $\alpha$ and is consistent.

\noindent (II) Under RL or CSR independence, the size of the above
one-sided tests are larger than $\alpha$ (i.e., the tests are
liberal in rejecting $H_o$) but consistent (in the sense that the
power goes to 1 as marginal sums tend to $\infty$ under the
alternatives).
\end{theorem}

\begin{theorem}
\label{thm:piel-chi2}
(I) Suppose the NNCT is based on a random sample of (base, NN) pairs.
The test for $H_a:\widetilde \pi_{11} \not= \widetilde \pi_{21}$ which
rejects $H_o:\,\widetilde \pi_{11} = \widetilde \pi_{21}$ for $\X_P^2>\chi^2_1(1-\al)$
with
$\X_P^2=\sum_{i=1}^2\sum_{j=1}^2\frac{(N_{ij}-\E[N_{ij}])^2}{\E[N_{ij}]}$
is consistent.

\noindent
(II) Under RL or CSR independence, the level of the test using $\X_P^2$ is larger than
$\al$ (i.e., it is liberal in rejecting these null patterns) but is consistent
in the sense of part (II) of Theorem \ref{thm:piel-directional}.
\end{theorem}

Next, we prove the consistency of Dixon's cell-specific and
overall tests of segregation.

\begin{theorem}
\label{thm:dixon-cell}
Under RL, Dixon's cell-specific test for cell $(i,j)$ in a NNCT denoted by $Z^D_{ij}$;
i.e., the test rejecting $H_o:\,
\pi_{ij}=\frac{n_{i}\,(n_i-1)}{n(n-1)}\I(i=j)+\frac{n_{i}\,n_j}{n(n-1)}\I(i\not=j)$
(i.e., RL ) against the two-sided (and one-sided
alternatives) for $|Z^D_{ij}|>z(1-\al/2)$ (and $Z^D_{ij}> z(1-\al)$
or $Z^D_{ij} < z(\al)$) with
$Z^D_{ij}=\frac{N_{ij}-\E[N_{ij}]}{\sqrt{\Var[N_{ij}]}}$ is of size
$\al$ and is consistent.
Under CSR independence, 
$Z^D_{ij}$ is consistent conditional on $Q$ and $R$.
\end{theorem}

\begin{theorem}
\label{thm:dixon-overall}
Under RL, Dixon's overall test of segregation; i.e., the test rejecting
$H_o:\,\pi_{ij}=\frac{n_{i}\,(n_i-1)}{n(n-1)}\I(i=j)+\frac{n_{i}\,n_j}{n(n-1)}\I(i\not=j)
\text{ for all } i,j \in \{1,2\}$ (i.e., RL)
against the alternative
$H_a:\,\pi_{ij}\not=\frac{n_{i}\,(n_i-1)}{n(n-1)}\I(i=j)+\frac{n_{i}\,n_j}{n(n-1)}\I(i\not=j)
\text{ for some } i,j \in \{1,2\}$ for $\X_D^2>\chi^2_{2}(1-\al)$
with
$\X_D^2=(\mathbf{N}-\E[\mathbf{N}])'\Sigma^{-}(\mathbf{N}-\E[\mathbf{N}])$
is of size $\al$ and is consistent.
Under CSR independence, 
$\X_D^2$ is consistent conditional on $Q$ and $R$.
\end{theorem}

\section{Monte Carlo Simulation Analysis}
\label{sec:Monte-Carlo}
Pielou's test of independence and Dixon's overall test of
segregation are not testing the same null pattern,
so we can not compare the power of the tests under
either segregation or association alternatives
and we only implement Monte Carlo simulations to evaluate
the finite sample performance of the tests in terms of empirical size.
For the null case,
we simulate the RL, CSR independence patterns,
and independence of the rows in the NNCTs,
with two classes labelled as $X$ and $Y$ with sizes $n_1$ and $n_2$, respectively.

\subsection{Empirical Significance Levels of the NNCT-Tests under RL}
\label{sec:Monte-Carlo-RL}
Under RL, we consider four cases.
In RL Case (1) we use the locations of the trees in the swamp tree data
(see Figure \ref{fig:Swamp} and \cite{dixon:1994}) as the fixed points,
and randomly assign $n_1=182$ points as $X$ and $n_2=91$ points as $Y$ points.
In each of the other RL cases, we first determine the fixed locations
of points for which class labels are to be assigned randomly.
Then we apply the RL procedure to these points for
respective sample size combinations as follows.

\noindent
\textbf{RL Case (2):}
First, we generate $n=n_1+n_2$ points iid $\U((0,1) \times (0,1))$,
the uniform distribution on the unit square,
for some combinations of $n_1,n_2 \in \{10,30,50,100\}$.
In each $(n_1,n_2)$ combination, the locations of these points
are taken to be the fixed locations
for which we assign the class labels randomly.
For each sample size combination $(n_1,n_2)$,
we randomly choose $n_1$ points (without replacement) and label them
as $X$ and the remaining $n_2$ points as $Y$ points,
and repeat the RL procedure $N_{mc}=10000$ times.
At each Monte Carlo replication, we compute the NNCT-tests.
Out of these 10000 samples the number of significant outcomes by each test is recorded.
The nominal significance level used in all these tests is $\alpha=.05$.
The empirical sizes are calculated as
the ratio of number of significant results to the number of Monte
Carlo replications, $N_{mc}$.
That is, for example empirical size for Dixon's overall test for $(10,10)$, denoted by $\ah_{D}$,
is calculated as $\ah_{D}:=\sum_{i=1}^{N_{mc}} \I(\X^2_{D,i} \ge \chi^2_2(.05))$
where $\X^2_{D,i}$ is the value of Dixon's overall test statistic for iteration $i$,
$\chi^2_2(.05)$ is the $95^{th}$ percentile of $\chi^2_2$ distribution,
and $\I(\cdot)$ is the indicator function.

\noindent
\textbf{RL Case (3):}
We generate $n_1$ points iid $\U((0,2/3) \times (0,2/3))$ and
$n_2$ points iid $\U((1/3,1) \times (1/3,1))$
for some combinations of $n_1,n_2 \in \{10,30,50,100\}$.
The locations of these points are taken to be the fixed locations
for which we assign the class labels randomly.
The RL procedure is applied to these fixed points $N_{mc}=10000$
times for each sample size combination and
the empirical sizes for the tests are calculated similarly as in RL Case (2).

\noindent
\textbf{RL Case (4):}
We generate $n_1$ points iid $\U((0,1) \times (0,1))$
and $n_2$ points iid $\U((2,3) \times (0,1))$
for some combinations of $n_1,n_2 \in \{10,30,50,100\}$.
The RL procedure is applied and the empirical sizes for the tests
are calculated as in the previous RL Cases.

\begin{figure}[ht]
\centering
\rotatebox{-90}{ \resizebox{2. in}{!}{\includegraphics{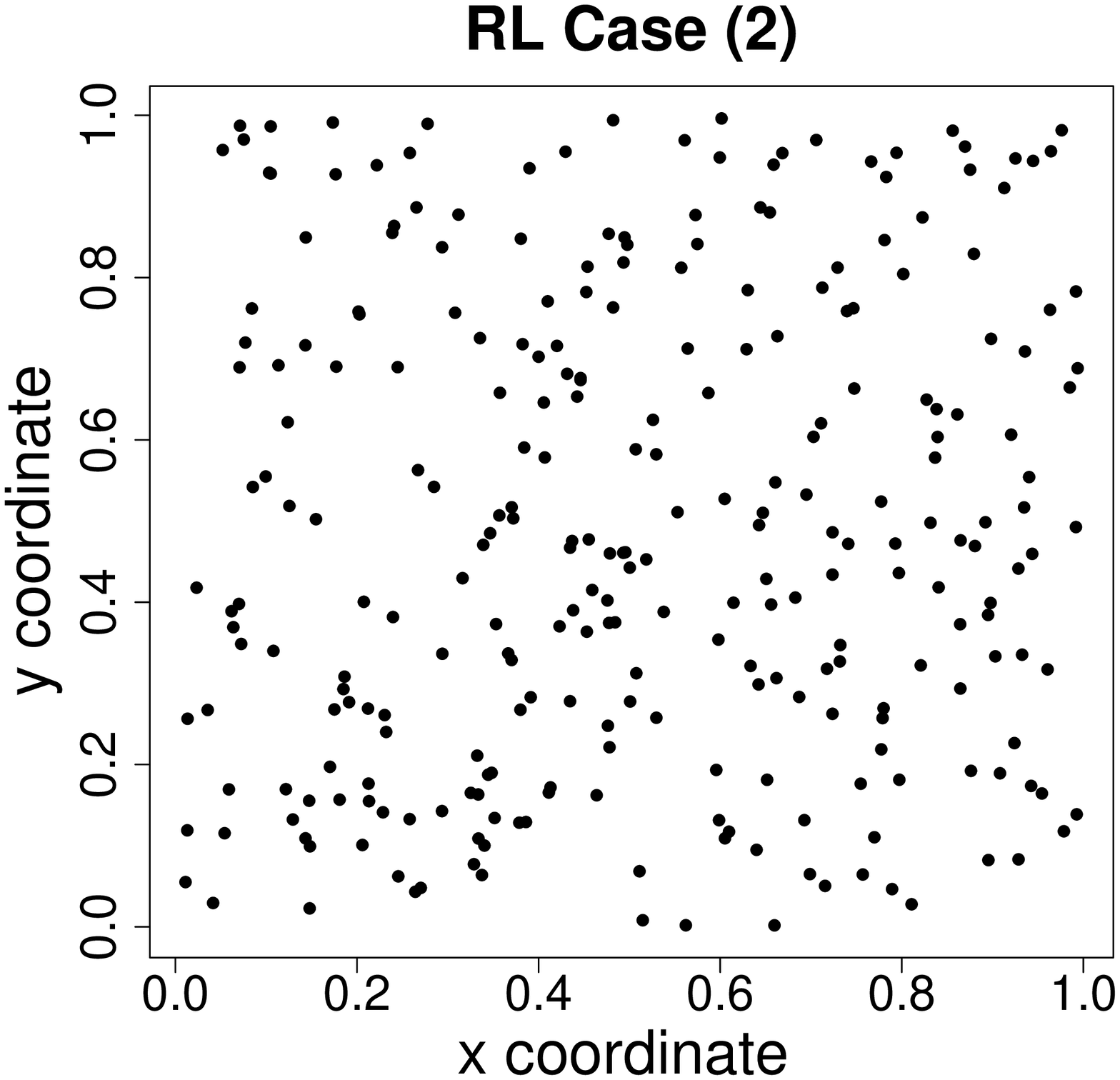} }}
\rotatebox{-90}{ \resizebox{2. in}{!}{\includegraphics{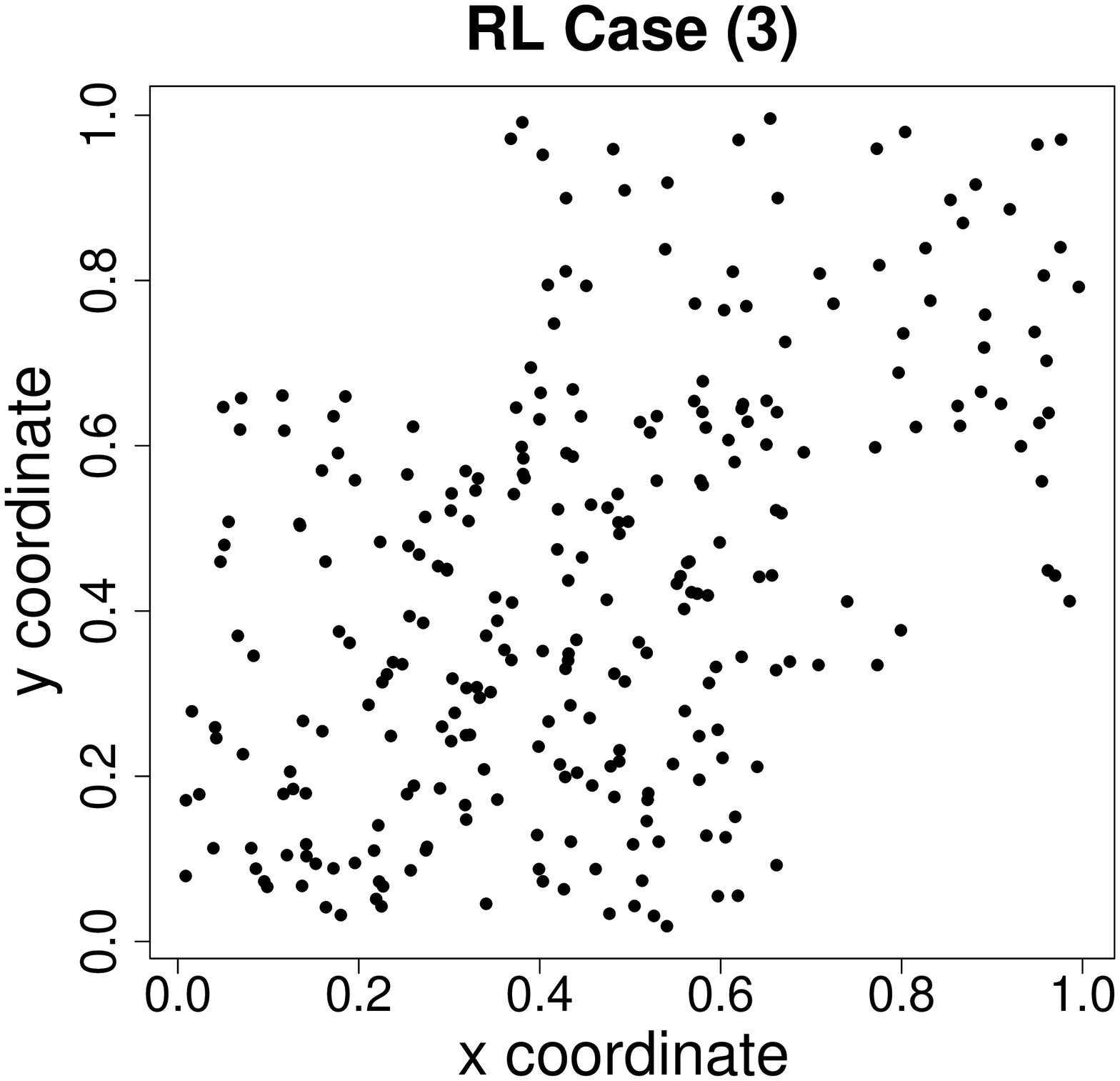} }}
\rotatebox{-90}{ \resizebox{2. in}{!}{\includegraphics{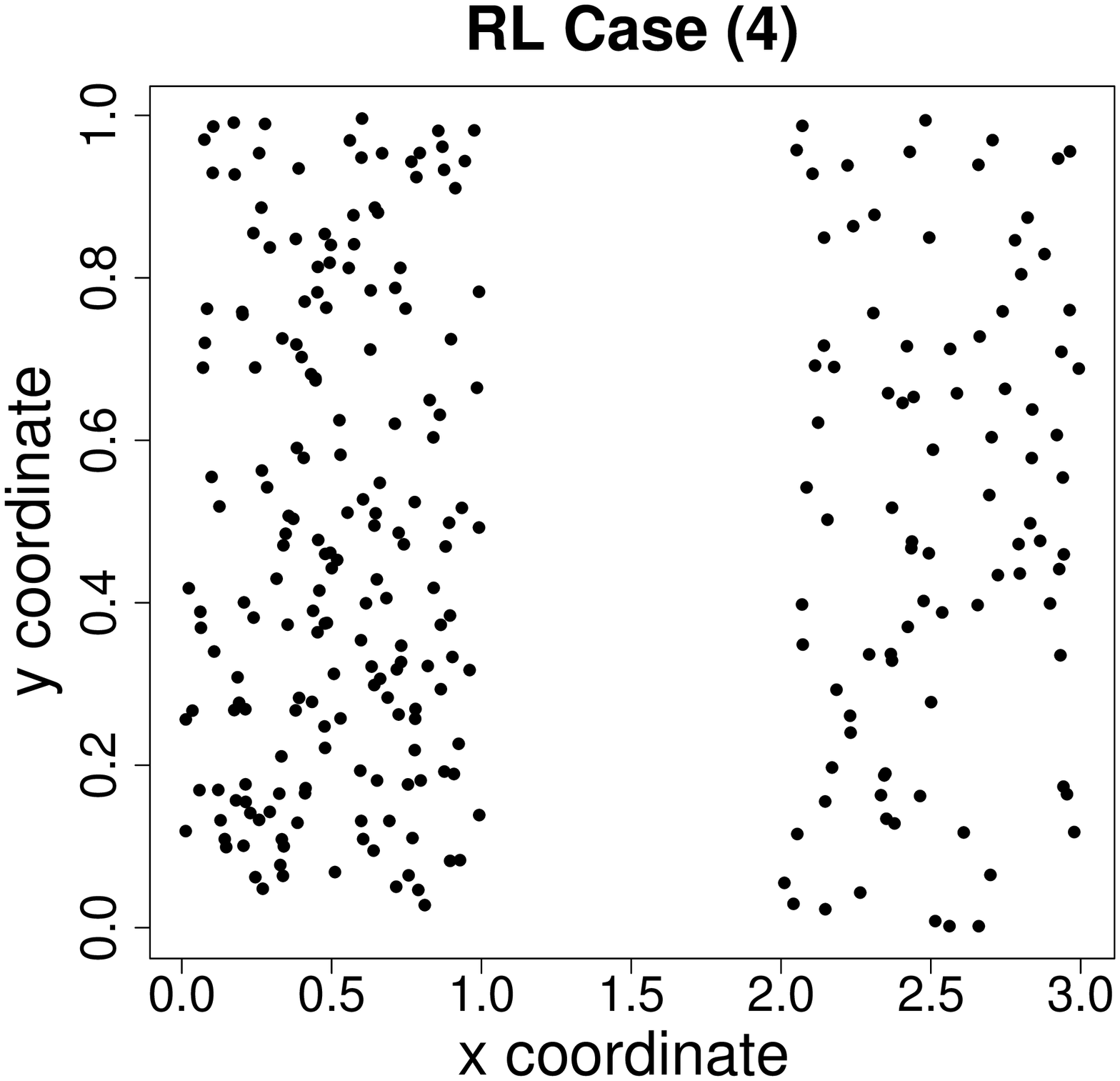} }}
 \caption{
\label{fig:RL-cases}
The fixed locations of points for which RL procedure is applied
for RL Cases (2)-(4) with $n=200$ (for the case with $n_1=n_2=100$) in the two-class case.
Notice that $x$-axis for RL Case (4) is differently scaled than others.
}
\end{figure}

The locations for which the RL procedure is applied in RL Cases (2)-(4) are plotted
in Figure \ref{fig:RL-cases} for $n_1=n_2=100$.
Although there are many possibilities for the allocation of
points to which RL procedure can be applied,
we only chose the locations of trees in a real life data set
and three generic cases.
Observe that in RL Case (2), the allocation of the points are a realization of a homogeneous
Poisson process in the unit square;
in RL Case (3) the points are a realization of two overlapping clusters;
in RL Case (4) the points are a realization of two disjoint clusters.

The empirical significance levels are presented in Table \ref{tab:overall-null-RL},
where $\ah_{i,i}^D$ is the empirical
significance level for cell $(i,i)$ with $i \in \{1,2\}$,
$\ah_R$ and $\ah_L$ are the estimated empirical
significance levels for the right- and left-sided versions of Pielou's test, respectively
(see Equation \eqref{eqn:2x2-Zform}),
$\ah_{P}$ and $\ah_{PY}$ are for Pielou's overall test of
segregation without and with Yates' correction, respectively,
$\ah_{D}$ is for Dixon's overall test of segregation
with $n_1,n_2 \in \{10,30,50,100\}$ and $N_{mc}=10000$.
Notice that among the cell-specific tests
only $\ah_{1,1}^D$ and $\ah_{2,2}^D$ are presented in Table \ref{tab:null-CSR},
since $N_{12}=n_1-N_{11}$ and $N_{21}=n_2-N_{22}$ which implies
$\ah_{1,1}^D=\ah_{1,2}^D$ and $\ah_{2,1}^D=\ah_{2,2}^D$ in the two-class case.
The empirical sizes significantly smaller (larger) than .05 are
marked with $^c$ ($^{\ell}$),
which indicate that the corresponding test is conservative (liberal).
The asymptotic normal approximation
to proportions is used in determining the significance of the
deviations of the empirical size estimates from the nominal level of .05.
For these proportion  tests, we also use $\alpha=.05$ to test against
empirical size being equal to .05.
With $N_{mc}=10000$, empirical
sizes less (greater) than .0464 (.0536) are deemed conservative (liberal) at $\alpha=.05$ level.
Observe that Dixon's cell-specific tests
are slightly liberal or conservative or about the desired significance levels in
rejecting $H_o:$ RL  when $n_1,n_2 \ge 30$.
When $n_i \le 10$ for $i=1$ or 2, then
Dixon's cell-specific tests tend to be conservative if $n_1 \not= n_2$
and liberal otherwise.
Notice also that when $n_i \le 10$ for $i=1$ or 2 and $n_1 \not= n_2$
Dixon's cell-specific test is more conservative for cell $(1,1)$ which corresponds
to the class with smaller size (i.e., class $X$) compared to class $Y$.
On the other hand, Pielou's overall test and the right sided version of Pielou's test
are extremely liberal for all sample size combinations;
left sided version of Pielou's test is extremely liberal for
all sample size combinations except for $(10,50)$.
Furthermore, Pielou's test with Yates' correction
is liberal when $\min(n_1,n_2) \ge 30$, conservative for $(10,50)$
and liberal or about the nominal level for other sample size combinations.
Notice also that, $\ah_{PY}$ values are significantly smaller
(based on the tests of equality of the proportions for two populations)
compared to $\ah_{P}$ values.
Dixon's overall test of segregation
tends to be conservative for $(10,10)$ and $(10,30)$
and is about the desired nominal level for most
of the other sample size combinations.
These results suggest that under RL,
Dixon's tests (especially the overall test) are appropriate,
but Pielou's tests are not.

\subsection{Empirical Significance Levels of the NNCT-Tests under CSR Independence}
\label{sec:Monte-Carlo-CSR}
Under CSR independence, at each of $N_{mc}=10000$ replicates,
we generate points iid from $\U((0,1)\times (0,1))$,
for some combinations of $n_1,n_2 \in \{10,30,50,100\}$.
Let $X=\{X_1,\ldots,X_{n_1}\}$ be the set of class 1 points and
$Y=\{Y_1,\ldots,Y_{n_2}\}$ be the set of class 2 points.

We present the empirical significance
levels for Pielou's tests and Dixon's tests in Table \ref{tab:null-CSR},
where $\ah_{i,i}^{D,qr}$ is Dixon's cell-specific test for cell $(i,i)$
and $\ah_{D,qr}$ is Dixon's overall test with $Q$ and $R$ are replaced with their (empirical) expected values
and notations for the other tests are as in Section \ref{sec:Monte-Carlo-RL}.
The empirical sizes are calculated for some combinations of $n_1,n_2 \in \{10,30,50,100\}$ and $N_{mc}=10000$.
Observe that Dixon's cell-specific tests tend to be slightly liberal or conservative
or about the desired significance level for most sample size combinations.
In particular, they have about the nominal level for $n_1=n_2 \ge 30$,
and are extremely conservative for the smaller class when $n_i=10$ for one of $i=1,2$
(see, e.g, $\ah_{1,1}^D$ for $(n_1,n_2)=(10,50)$ and $\ah_{2,2}^D$ for $(n_1,n_2)=(50,10)$).
The QR-adjusted versions of cell-specific tests tend to have different
sizes than the uncorrected versions when $n_i \le 30$ for $i=1$ or 2.
For larger samples, QR-adjustment does not improve the sizes
compared to the uncorrected ones.
Pielou's overall test and one-sided versions
are extremely liberal for all sample sizes
(however, notice that Pielou's overall and left-sided tests are
least liberal for $(n_1,n_2)=(10,50)$ and $(n_1,n_2)=(50,10)$).
Pielou's test with Yates' correction is at the nominal level
for $(n_1,n_2)=(30,10)$, conservative for $(n_1,n_2)=(10,50)$ and $(n_1,n_2)=(50,10)$, and
liberal for other sample size combinations.
Notice also that $\ah_{PY}$ values are significantly smaller
than $\ah_{P}$ values (i.e., Yates' correction significantly reduces
the empirical size for Pielou's overall test).
As for Dixon's overall test,
it tends to be conservative for small sample size combinations,
and is about the desired level for most large sample size combinations.
As in the cell-specific tests, QR-adjustment
does not improve on the uncorrected versions.
For more detail on QR-adjustment for NNCT-tests, see (\cite{ceyhan:2008}).
Hence \emph{in the following sections,
we only provide the uncorrected versions of Dixon's tests}.

\begin{remark}
\label{rem:prop-agree}
\textbf{Proportion of Agreement between Pielou's and Dixon's Overall Tests:}
At each sample size combination under the RL Cases (2)-(5) and CSR independence,
we also record the number of times both Pielou's and Dixon's overall tests
simultaneously yield significant results at $\alpha=.05$.
The ratio of number of significant results by both tests
to the number of Monte Carlo replications, $N_{mc}$,
is the proportion of agreement between the tests
in rejecting the particular null pattern.
That is, for example the proportion of agreement between Pielou's and Dixon's overall tests
denoted by $\ah_{P,D}$ for $(n_1,n_2)=(10,10)$ under RL Case (2)
is calculated as $\ah_{P,D}:=\sum_{i=1}^{N_{mc}} \I(\X^2_{P,i} \ge \chi^2_1(0.95))\I(\X^2_{D,i} \ge \chi^2_2(0.95))$
where $\X^2_{P,i}$ is the value of Pielou's overall test statistic for iteration $i$.
The estimates of the proportion of agreement values are presented in Table \ref{tab:prop-agree}.
Observe that $\ah_{P,D}$ values are significantly smaller than $\min(\ah_{P},\ah_{D})=\ah_{D}$
at each sample size combination under each null case.
This supports the discussion in the first paragraph of Section \ref{sec:consistency};
that is, the rejection regions (hence acceptance regions) for both tests are
significantly different and neither one is uniformly superior to the other.
$\square$
\end{remark}

\subsection{Empirical Significance Levels of the Tests under
Independence of Rows and Cell Counts in NNCTs}
\label{sec:Monte-Carlo-Independence}
For the independence of rows and cell counts in the NNCTs,
we consider two cases: overall multinomial and row-wise binomial frameworks.
In the overall multinomial case,
we generate all four cell counts using multinomial distribution,
$\mathscr M(n,\widetilde \pi_{11},\widetilde \pi_{12},\widetilde \pi_{21},\widetilde \pi_{22})$,
with $\widetilde \pi_{11}=\widetilde \pi_{21}=\frac{n_1}{2(n_1+n_2)}$ and
$\widetilde \pi_{12}=\widetilde \pi_{22}=\frac{n_2}{2(n_1+n_2)}$.
The NNCT constructed in such a way is (approximately) equivalent to one based
on a random sample of (base, NN) pairs.
In the row-wise binomial case, we generate the two cell counts in each row
using binomial distribution,
$N_{11}\sim \BIN(n_1,n_1/(n_1+n_2))$, and
$N_{21}\sim \BIN(n_2,n_1/(n_1+n_2))$.
The NNCT constructed in this way is also equivalent to one based
on a random sample of (base, NN) pairs.

For such NNCTs, we can only compute Pielou's test and
the one-sided versions, but not Dixon's tests,
since Dixon's tests require more information on the NN structure
in the spatial distribution of the points, e.g., the quantities
such as $Q$ and $R$, which are not available in this case.
That is, in any spatial allocation of points, the NN relations of
each point is dependent on the the relations of neighboring points,
which in turn implies that
it is not realistic to have a random sample of (base, NN) pairs in practice.

In Table \ref{tab:null-indep},
we present the empirical significance levels for all tests except Dixon's tests
under the independence of cells and rows case.
Observe that Pielou's test with Yates' correction
is extremely conservative under both of the frameworks.
On the other hand Pielou's one-sided tests and Pielou's test without Yates' correction
have about the desired nominal level.

\begin{remark}
\label{rem:main-result-emp-size}
\textbf{Main Result of Monte Carlo Simulations for Empirical Sizes:}
Based on the simulation results under the CSR of the points,
we recommend the disuse of Pielou's test in practice,
as it is extremely liberal, hence it might give false alarms when the pattern
is actually not significantly different from RL or CSR independence.
Moreover Yates' correction does not seem to fix the problems with Pielou's test,
since the problems are not caused by the discrete nature of the cell counts.
However, Yates' correction seems to improve the performance of Pielou's test,
in the sense that empirical size of Pielou's test with Yates' correction
gets closer to the nominal level compared to the uncorrected one.
Even Dixon's tests fail to
have the desired level when at least one sample size is small so that
the cell count(s) in the corresponding NNCT have a high probability of being $\le 5$.
This usually corresponds to the case that at least
one sample size is $\leq 10$ or the sample sizes (i.e., relative abundances) are very different
in our simulation study.
When sample sizes are small (hence the corresponding NNCT cell counts are $\leq 5$),
the asymptotic approximation of Dixon's tests is not appropriate.
So \cite{dixon:1994} recommends Monte Carlo randomization
for his test when some NNCT cell count(s) are $\le 5$ under RL.
We concur with the same recommendation for the RL pattern
and extend this recommendation for CSR independence.
In fact, this recommendation is also partly consistent with
the inapplicability of asymptotic results for contingency tables in general
(not just for NNCTs) when cell counts are too small.
In general contingency tables,
the chi-squared approximation seems to be valid
in most cases if all expected cell counts are larger than 0.5
and at least half are greater than 1.0 (\cite{conover:1999}).
On the other hand, \cite{cochran:1952} states
that the approximation may be poor if any expected
cell count is less than 1 or if more than 20 \%
of the expected cell counts are less than 5.
$\square$
\end{remark}

\section{Edge Correction for the CSR Independence Pattern}
\label{sec:edge-correct}
In this section, we investigate the edge or boundary effects on the
NNCT-tests used under CSR independence.
Edge effects arise because CSR independence assumes an unbounded region,
which is not the case in practical situations.
Edge effects on spatial pattern analysis and various correction
methods are discussed extensively in spatial pattern literature
(\cite{clark:1954} and \cite{cressie:1993}).
However, the effectiveness of edge
correction depends on the type of the statistic used
(see, e.g., \cite{yamada:2003}).
For example, when the study region is rectangular, the edge
effects can be minimized by including a \emph{buffer zone} or \emph{area} around the rectangle,
or alternatively, the rectangular region is transformed into a
\emph{torus} (\cite{dixon:EncycEnv2002}).
In literature, buffer area
is also referred to as {\em guard area} (\cite{yamada:2003}).
The general idea is the same for buffer zone and toroidal edge corrections,
but they are implemented in different ways.
In buffer zone correction we assume the properties of the process are the same,
even if we continue into the buffer area.
In toroidal correction, the process is \emph{exactly} the same outside of the study area.

Without any edge correction the cell counts in a NNCT
can be written as
$$N_{ij}=\sum_{k=1}^n\sum_{l=1}^n w_{kl}S_{kl}\I(l\not=k)=
\sum_{k,l=1}^n w_{kl}S_{kl}\I(l\not=k)$$
where $S_{kl}$ is 1 if point $l$ is of class $j$ and point $k$ is of class $i$,
and 0 otherwise; $w_{kl}$ is 1 if point $l$ is the NN of point $k$, and 0 otherwise.
The quantities $Q$ and $R$ can be written as
$$Q=\sum_{m,k,l=1}^n w_{kl}w_{ml}\I(m\not=k\not=l)~~~
\text{ and }~~~
R=\sum_{k,l=1}^n w_{kl}w_{lk}\I(l\not=k),$$
with the understanding that $\I(m\not=k\not=l)=\I(m\not=k)\I(m\not=l)\I(k\not=l)$.

\subsection{Buffer Zone Correction for the CSR Independence Pattern}
\label{sec:buffer-edge-correct}
In the \emph{buffer zone correction method}, a guard
area is selected inside or outside the study region and the points
in the guard area are used only as destinations
(not the base points) in NN relations.
In the NN pair $(U,V)$, point $U$ is the \emph{base point},
and point $V$ is the \emph{destination point}.
When the buffer area is sufficiently large,
the edge effects can be completely eliminated,
but this is a wasteful procedure, because the large buffer area may
contain many observations.

Let $R_O$ be the original study area,
$R_B$ be the outer buffer area,
and $R_b$ be the inner buffer area
and let $n$ be the number of points that fall in $R_O$.
In the outer buffer zone correction,
let $n_B$ be the number of points that fall in $R_O \cup R_B$,
and points with indices $1,2,\ldots,n$ lie in $R_O$,
and points with indices $(n+1),(n+2),\ldots,n_B$ lie in $R_B$.
With the outer buffer zone correction, the NNCT cell counts are
$$N_{ij}=\sum_{k=1}^n\sum_{l=1}^{n_B} w_{kl}S_{kl}\I(k\not=l).$$
Furthermore, the quantities $Q$ and $R$ are also modified as follows:
$$
Q=\sum_{m,k=1}^{n_B}\sum_{l=1}^n w_{kl}w_{ml}\I(m\not=k\not=l)
\text{ and }
R=\sum_{k,l=1}^{n_B} w_{kl}w_{lk}\I(k\not=l)-
\sum_{k,l=(n+1)}^{n_B} w_{kl}w_{lk}\I(k\not=l).
$$

In the inner buffer zone correction,
let $n_b$ be the number of points in the inner buffer area $R_b$,
and points with indices $1,2,\ldots,(n-n_b)$ lie in $R_O \setminus R_b$,
and points with indices $(n-n_b+1),(n-n_b+2),\ldots,n$ lie in $R_b$.
Then, with the inner buffer zone correction, the NNCT cell counts are
$$N_{ij}=\sum_{k=1}^{n-n_b}\sum_{l=1}^{n} w_{kl}S_{kl}\I(k\not=l).$$
Furthermore, the quantities $Q$ and $R$ are
$$
Q=\sum_{m,k=1}^{n}\sum_{l=1}^{n-n_b} w_{kl}w_{ml}\I(m\not=k\not=l)
\text{ and }
R=\sum_{k,l=1}^{n} w_{kl}w_{lk}\I(k\not=l)-
\sum_{k,l=(n-n_b+1)}^{n_B} w_{kl}w_{lk}\I(k\not=l).
$$

In our simulation study,
we consider the edge correction by outer buffer zone correction only.
For CSR independence, we generate points iid from $\U((-.5,1.5)\times
(-.5,1.5))$ until there are $n_1$ class $X$ points and $n_2$ class
$Y$ points in the unit square for some combinations of $n_1,n_2 \in \{10,30,50,100\}$.
We repeat this procedure $N_{mc}=10000$ times for
each $n_1,n_2$ combination.
The corresponding empirical significance
levels are provided in Table \ref{tab:overall-buff}.
Observe that compared to the uncorrected sizes,
with the (outer) buffer zone edge correction,
the empirical sizes of the right-sided version of Pielou's test
do not significantly change;
empirical sizes of the left-sided version of Pielou's test
significantly decrease for smaller samples,
do not significantly change for other sample size combinations;
empirical sizes of Pielou's test with Yates' correction
do not significantly change (except for $(10,10)$);
empirical sizes of Dixon's cell-specific tests do not change for most sample size combinations.
On the other hand, for Dixon's test,
the empirical sizes significantly increase to become liberal for $n_1+n_2 \le 80$.
Furthermore Pielou's overall and one-sided tests
still tend to be liberal with the (outer) buffer zone correction;
Pielou's test with Yates' correction is liberal when $n_1+n_2 \ge 40$.
Dixon's cell-specific tests become about the desired level when $n_i \ge 50$
for both $i=1,2$,
the sizes do not change or improve in one direction for smaller samples.

\begin{remark}
\label{rem:inner-vs-outer-buffer}
\textbf{Inner versus Outer Buffer Zone Correction:}
The main difference between inner and outer buffer zone correction
is the time of the selection of the buffer zone.
In the outer buffer zone correction,
a larger region than the intended region of interest is selected
prior to recording the observations;
while in the inner buffer zone correction
some part of the original study region is designated as the buffer zone
after the data is collected.
Hence, theoretically the inner and outer buffer zones behave similarly.
Indeed, $R_O \setminus R_b$ in the inner buffer zone correction
acts like the original region $R_O$ of the outer buffer zone correction,
and likewise $R_b$ in inner buffer zone correction acts like
$R_B$ of the outer buffer zone correction.
Thus, we only simulate the outer buffer zone correction,
as it can also be equivalently viewed as the inner buffer zone correction.
That is, the square $(-.5,1.5)\times (-.5,1.5)$ can be viewed as $R_O$
and $[(-.5,1.5)\times (-.5,1.5)]\setminus [(0,1)\times(0,1)]$ can be viewed as $R_b$.
$\square$
\end{remark}

\subsection{Toroidal Edge Correction for the CSR Independence Pattern}
\label{sec:toroidal-edge-correct}
In the \emph{toroidal edge correction},
the original area is surrounded by eight copies of the original
study area and the points in these additional copies are used only
as destination points. 
For the toroidal edge correction, clusters around
the boundaries might cause bias.
Moreover, while toroidal correction
applies only to rectangular study regions, buffer zone correction
applies to any type of study region (see, (\cite{yamada:2003}).

Let $R_O$ be the original study area, $R_T$ be the eight copies appended to $R_O$
so as to obtain NN structure for the points in $R_O$ as if $R_O$
is part of a torus.
For the toroidal correction,
let $n_T$ be the number of points that fall in $R_O \cup R_T$,
and points with indices $1,2,\ldots,n$ lie in $R_O$,
and points with indices $(n+1),(n+2),\ldots,n_T$ lie in the toroidal area $R_T$.
With the toroidal correction, the NNCT cell counts are
$$N_{ij}=\sum_{k=1}^n\sum_{l=1}^{n_T} w_{kl}S_{kl}\I(k\not=l).$$
Furthermore, the quantities $Q$ and $R$ are also modified as follows:
$$
Q=\sum_{m,k=1}^{n_T} \sum_{l=1}^n w_{kl}w_{ml}\I(m\not=k\not=l),
\text{ and }
R=\sum_{k,l=1}^{n_T} w_{kl}w_{lk}\I(k\not=l)-
\sum_{k,l=(n+1)}^{n_T} w_{kl}w_{lk}\I(k\not=l).
$$

For toroidal correction, under $H_o:$ CSR independence, we generate $n_1$
$X$-points and $n_2$ $Y$-points iid from $\U((0,1)\times (0,1))$ for
some combinations of $n_1,n_2 \in \{10,30,50,100\}$.
We repeat this
procedure $N_{mc}=10000$ times for each $n_1,n_2$ combination.
The corresponding empirical significance levels are presented in Table \ref{tab:overall-tor}.
Observe that toroidal edge correction does
not significantly affect the empirical sizes of the NNCT-tests.

\begin{remark}
\label{rem:main-result-edge-correct}
\textbf{Main Result of Edge Correction Analysis:}
The Monte Carlo analysis in Section \ref{sec:edge-correct}
suggests that the empirical sizes of the
NNCT-tests are not affected by the toroidal edge correction
because in our Monte Carlo simulations,
we have generated the CSR independence pattern on the unit square.
Any clusters in a realization of CSR are due to chance and are equally
likely to occur anywhere, so the clusters are more likely to occur
away from the boundary of the region.
However, the (outer) buffer zone edge correction method
seems to have stronger influence on the tests compared to toroidal correction.
In particular, the empirical sizes of the Dixon's test
tend to significantly increase with buffer zone correction.
But for all other tests, buffer zone correction
does not change the sizes significantly for most sample size combinations.

This is in agreement with the findings of \cite{barot:1999}
which says NN methods only require a small buffer area around the study region.
A large buffer area does not help too much since one only
needs to be able to see far enough away from an event to find its NN.
Once the buffer area extends past the likely NN distances
(i.e., about the average NN distances),
it is not adding much helpful information for NNCTs.
Furthermore, since buffer (inner or outer) zone correction methods are wasteful,
and strongly depend on the size of the zone,
we do not recommend their use for NNCT-tests.
On the other hand, one can use toroidal edge correction,
but the gain might not be worth the effort.
$\square$
\end{remark}

\section{Examples}
\label{sec:examples}
We illustrate the tests on two example data sets:
Dixon's swamp tree data (\cite{dixon:EncycEnv2002}) and an artificial data set.
We present the corresponding NNCTs, test statistics,
and edge correction results for both examples.
Since an outer buffer zone is not provided in these examples,
inner buffer zone correction is the only type
of buffer zone correction we can apply.
Moreover, since the regions are rectangular,
we can also apply toroidal edge correction in these examples.

\subsection{Swamp Tree Data}
\label{sec:dixon-data}
\begin{figure}[ht]
\centering
\rotatebox{-90}{ \resizebox{3.0 in}{!}{
\includegraphics{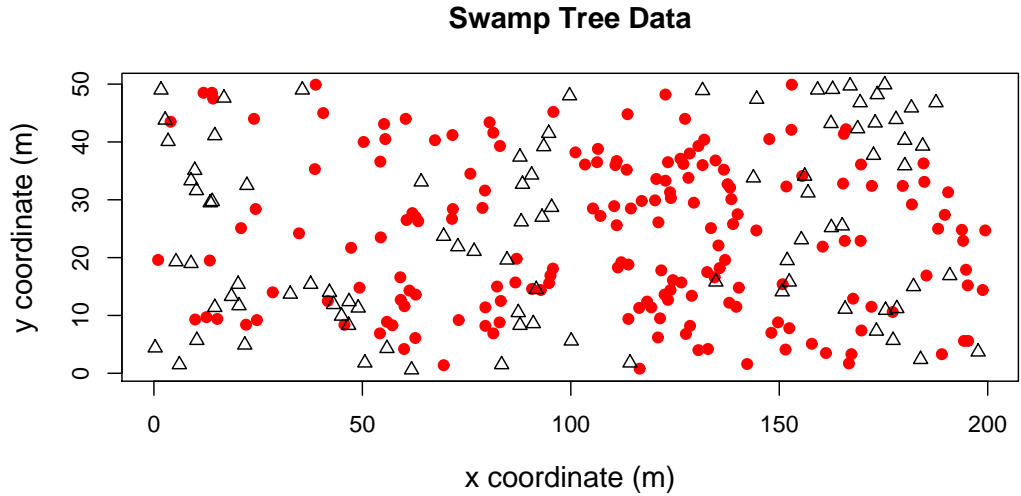} }}
\caption{\label{fig:Swamp}
The scatter plots of the locations of black gum trees (solid circles) and
bald cypress trees (triangles).}
\end{figure}

\cite{dixon:EncycEnv2002} illustrates NN-methods on
tree species in a 50 m by 200 m rectangular plot of hardwood swamp
in South Carolina, USA.
The plot contains trees from 13 different tree species,
of which we only consider the live trees from two
species, namely, black gums and bald cypresses.
If spatial interaction among the less frequent species were important,
a more detailed $12 \times 12$ NNCT-analysis should be performed.
For more detail on the data, see (\cite{dixon:EncycEnv2002}).
The locations of these trees in the study region are plotted in Figure
\ref{fig:Swamp}.

\cite{dixon:NNCTEco2002} applies his methodology for this data set assuming
the null pattern is the RL of tree species to the given locations.
But it is more reasonable to assume that the locations of
the tree species a priori result from different processes.
Hence the more appropriate null hypothesis would be the CSR independence pattern,
which implies that NNCT-test results are conditional ones.
The question of interest is whether the two tree species are
segregated, associated, or do not deviate from the CSR independence pattern.
The corresponding NNCT and the percentages are provided in Table \ref{tab:NNCT-example}.
The percentages for the cells are based on
the sample size of each species.
That is, for example 82 \% of black gums have NNs from black gums,
and remaining NNs of black gums are from bald cypresses.
The row and column percentages are marginal
percentages with respect to the total sample size.
The percentage values are also suggestive of segregation,
especially for black gum trees.

For the raw data (i.e., data not corrected for edge effects),
we find $Q=178$ and $R=156$.
The test statistics are provided in Table \ref{tab:example-test-stat},
where $\X^2_D$ stands for Dixon's overall segregation test,
$\X^2_P$ and $\X^2_{PY}$ for Pielou's test without and
with Yates' correction, respectively,
$Z_n$ for the directional $Z$-test.
The $p$-values are for the general alternative of deviation from CSR independence
for $\X^2_D$, $\X^2_P$, and $\X^2_{PY}$;
and for $Z_n$, the first $p$-value in the parenthesis
is for the association alternative,
while the second is for the segregation alternative.
Observe that all two-sided tests are significant, implying
significant deviation from CSR independence. 
The directional (one-sided) tests indicate that 
black gum trees and bald cypresses are significantly segregated.

Table \ref{tab:example-test-stat} also contains the $p$-values
when the edge correction methods are applied.
The toroidal correction does slightly change the test statistics,
but not the conclusions.
For the inner buffer zone correction,
we first estimate the density of the combined species, namely
$\widehat{\lambda}=(n_1+n_2)/A_r$ where $A_r$ is the area of the study region.
Let $W$ be the distance from a randomly chosen event to its NN,
then under CSR independence, $\E[W]=1\Big/\left(2\,\sqrt{\lambda} \right)$ and
$\Var[W]=(4-\pi)/(4\,\pi\,\lambda)$, where $\lambda$ is the intensity of
the point process (\cite{dixon:EncycEnv2002}).
So, we move the
boundaries inside the rectangular study area by
$\E[W]+k\,\sqrt{\Var[W]}=1\Big/\left(2\,\sqrt{\widehat{\lambda}}\right)
+k\,\sqrt{(4-\pi)\Big/\left(4\,\pi\,\widehat{\lambda}\right)}$ where $k$ determines
how much of the study region is regarded as the buffer zone.
We implement the inner buffer zone correction
with $k=0,1,2,3$, but only present the results for $k=0,1$,
since each case yields similar test statistics and the same conclusions.

\begin{figure}[t]
\centering
\rotatebox{-90}{ \resizebox{2 in}{!}{\includegraphics{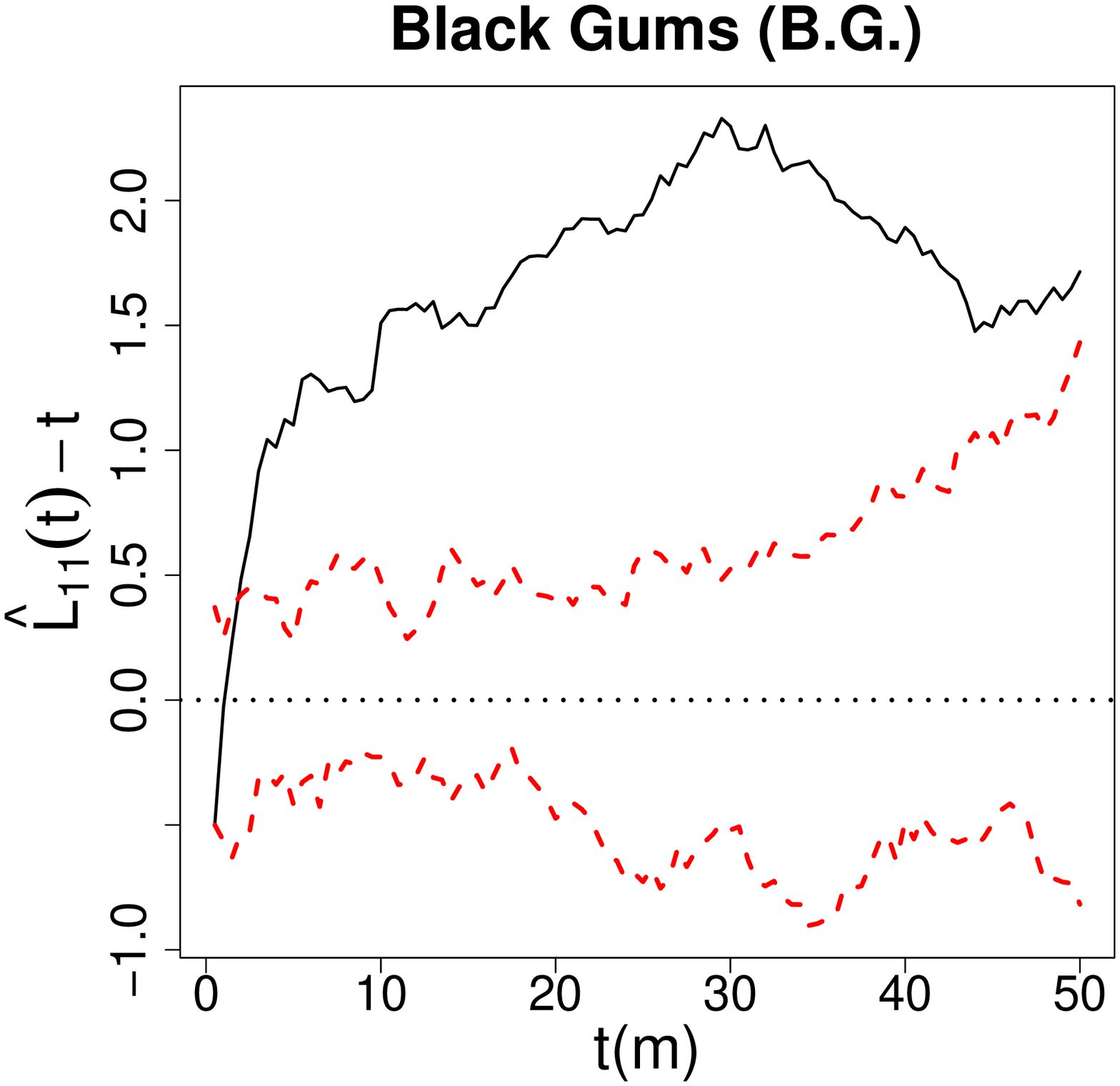} }}
\rotatebox{-90}{ \resizebox{2 in}{!}{\includegraphics{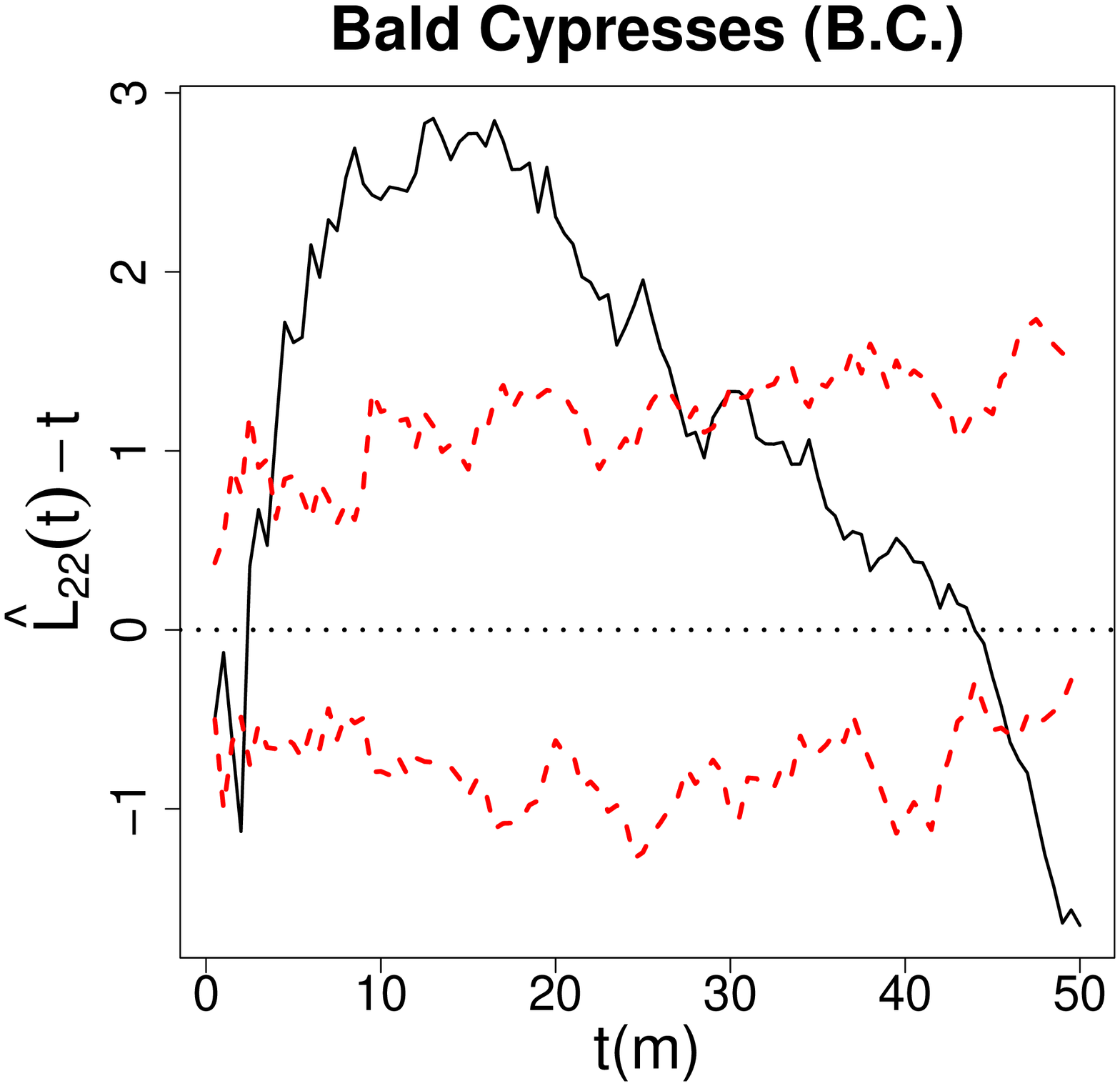} }}
\rotatebox{-90}{ \resizebox{2 in}{!}{\includegraphics{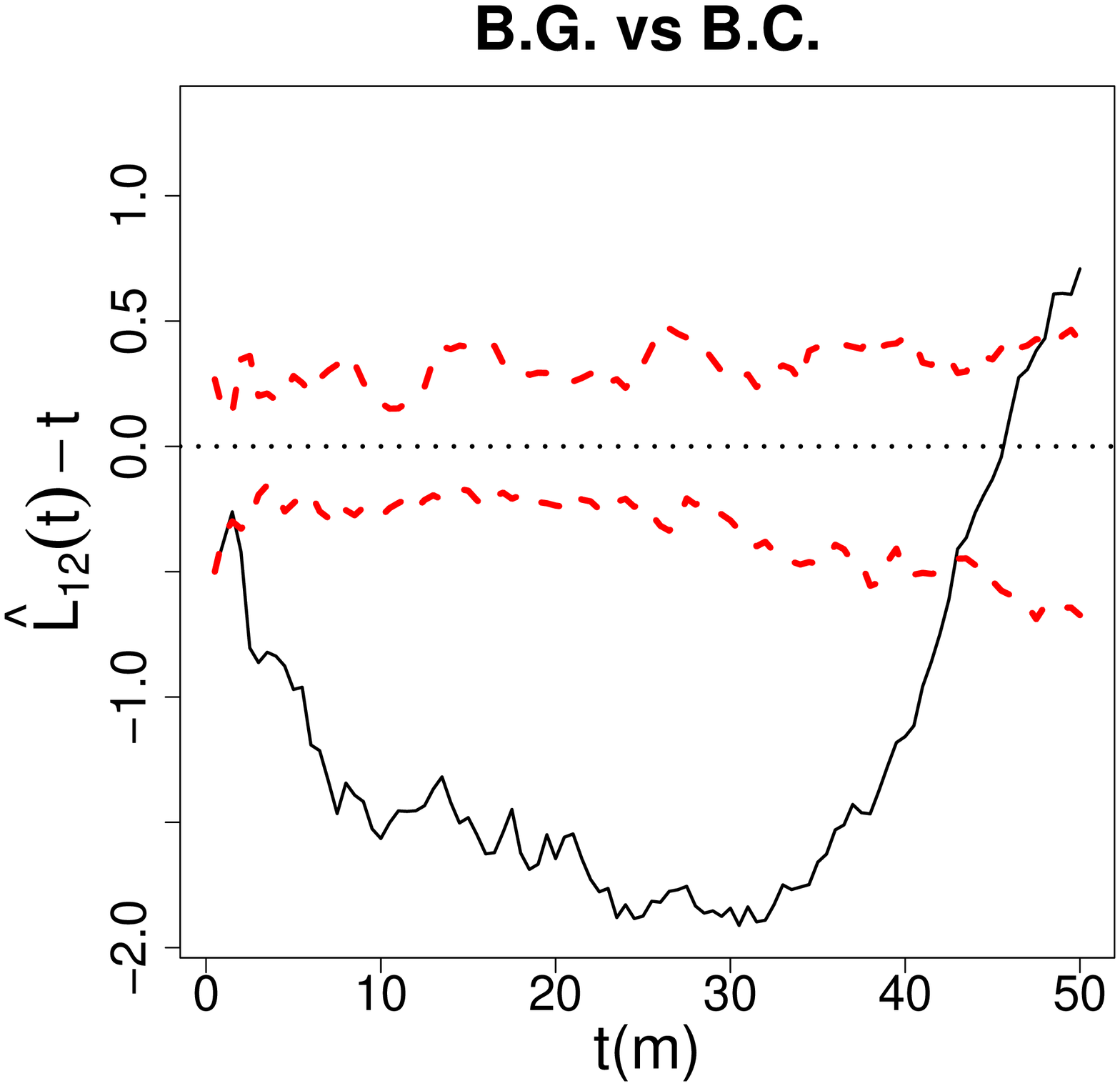} }}
\caption{
\label{fig:swamp-Liihat}
Second-order properties of swamp tree data.
Functions plotted are Ripley's univariate $L$-functions
$\widehat{L}_{ii}(t)-t$ for $i=1,2$,
and bivariate $L$-function $\widehat{L}_{12}(t)-t$
where $i=1$ for black gums and $i=2$ for bald cypresses.
The thick dashed lines around 0 are the upper and lower 95 \% confidence bounds for the
$L$-functions based on Monte Carlo simulation under the CSR independence pattern.
Notice also that vertical axes are differently scaled.}
\end{figure}

Based on the NNCT-tests, we conclude that tree species exhibit
significant deviation from the CSR independence pattern.
Considering Figure \ref{fig:Swamp}
and the corresponding NNCT in Table \ref{tab:NNCT-example},
this deviation is toward the segregation of the tree species.
However, the results of NNCT-tests pertain to small scale interaction,
i.e., at about the average NN distances.
To understand (possible) causes of the segregation
and the type and level of interaction between the tree species
at different scales (i.e., distances between the trees)
we also provide Ripley's univariate and bivariate
$L$-functions in Figure \ref{fig:swamp-Liihat},
where the spatial interaction is analyzed for distances up to 50 meters.

In Figure \ref{fig:swamp-Liihat},
we present the plots of Ripley's univariate $L$-function
$\widehat{L}_{ii}(t)-t$ for both species,
and bivariate $L$-function $\widehat{L}_{12}(t)$ for the pair of tree species.
Due to the symmetry of $L_{ij}(t)$, we omit $\widehat{L}_{21}(t)$.
We also present the upper and lower 95 \% confidence bounds for each $\widehat{L}_{ii}(t)-t$
and $\widehat{L}_{12}(t)-t$ under CSR independence.
Observe that black gums exhibit significant aggregation for distances $t>1$ m
(the $L_{11}(t)-t$ curve is above the upper confidence bound);
bald cypresses exhibit no deviation from CSR around $t \approx 5$ m,
then they exhibit significant spatial aggregation for $t$ up to 30 m,
then for larger distances ($t>45$ m) they exhibit spatial regularity.
Observe also that
black gums and bald cypresses are significantly segregated for distances
up to $t \approx 42$ m ($\widehat{L}_{12}(t)-t$ is below the lower confidence bound),
for larger distances their interaction
does not deviate significantly from CSR independence.
The segregation of the species might be due to different
levels and types of aggregation of the species in the study region.
Note also that average NN distance for swamp tree data is
$3.08 \pm 1.70$ (mean $\pm$ standard deviation) and results of
bivariate $L$-function and NNCT-analysis agree for distances around $t=3$ m.

\subsection{Artificial Data Set}
\label{sec:arti-data}
\begin{figure}[ht]
\centering
\rotatebox{-90}{ \resizebox{3. in}{!}{
\includegraphics{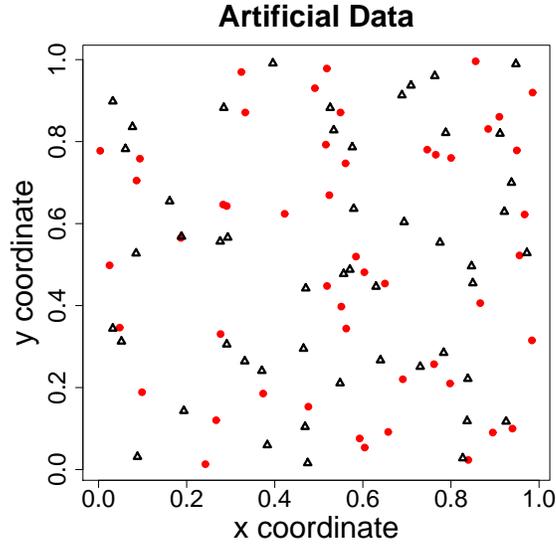} }}
 \caption{
 \label{fig:Arti}
The scatter plots of the locations of $X$ points (solid circles) and $Y$
points (triangles).}
\end{figure}

In the swamp tree example,
although the expected NNCT cell counts (not presented)
are different for Pielou's and Dixon's tests
and $p$-values for Dixon's tests are larger than others,
we have the same conclusion:
there is strong evidence for segregation of tree species.
Below, we present an artificial example,
a random sample of size 100 (with $50$ $X$-points and
$50$ $Y$-points uniformly generated on the unit square).
The question of interest is the spatial interaction between $X$ and $Y$ classes.
We plot the locations of the points
in Figure \ref{fig:Arti} and the corresponding NNCT together with
percentages are provided in Table \ref{tab:NNCT-example}.
Observe that the percentages are suggestive of mild segregation, with
equal degree for both classes.

The data is generated to resemble the CSR independence pattern,
so we assume the null pattern is CSR independence,
which implies that our inference will be a conditional one.
Observe that in Table \ref{tab:example-test-stat},
Pielou's tests are significant while Dixon's test are not,
which might be interpreted as evidence of deviation from CSR independence.
The graph in Figure \ref{fig:Arti} is not
suggestive of any such deviation from CSR independence, and the dependence
between NNCT cell counts might be confounding the conclusions based on Pielou's tests.
With toroidal correction, all $p$-values increase,
but only for Pielou's test with
Yates' correction becomes insignificant after the correction.
With buffer zone correction
with $k=0$, all $p$-values get to be insignificant,
except for the one-sided test for segregation.
Furthermore, with $k=1$, the changes
in $p$-values are more dramatic.
We also have similar changes with $k=2,3$ (not presented).
Thus, inner buffer zone edge correction,
might make a big difference
if the pattern is a close call between CSR independence and
segregation/association.
That is, if a segregation test has a $p$-value about .05,
when a subregion is reserved as the inner buffer zone,
either we might have a pattern different from CSR independence
in the area for the base points (i.e., $R_O \setminus R_b$)
or after the loss of data in the buffer zone the power of the tests might decrease.
We also point out that, inner and outer buffer zone correction
methods are not recommended in literature
for spatial pattern analysis with, e.g., Ripley's $K$-function (\cite{yamada:2003}),
and we concur with this suggestion for NNCT-tests.

\begin{figure}[t]
\centering
\rotatebox{-90}{ \resizebox{2 in}{!}{\includegraphics{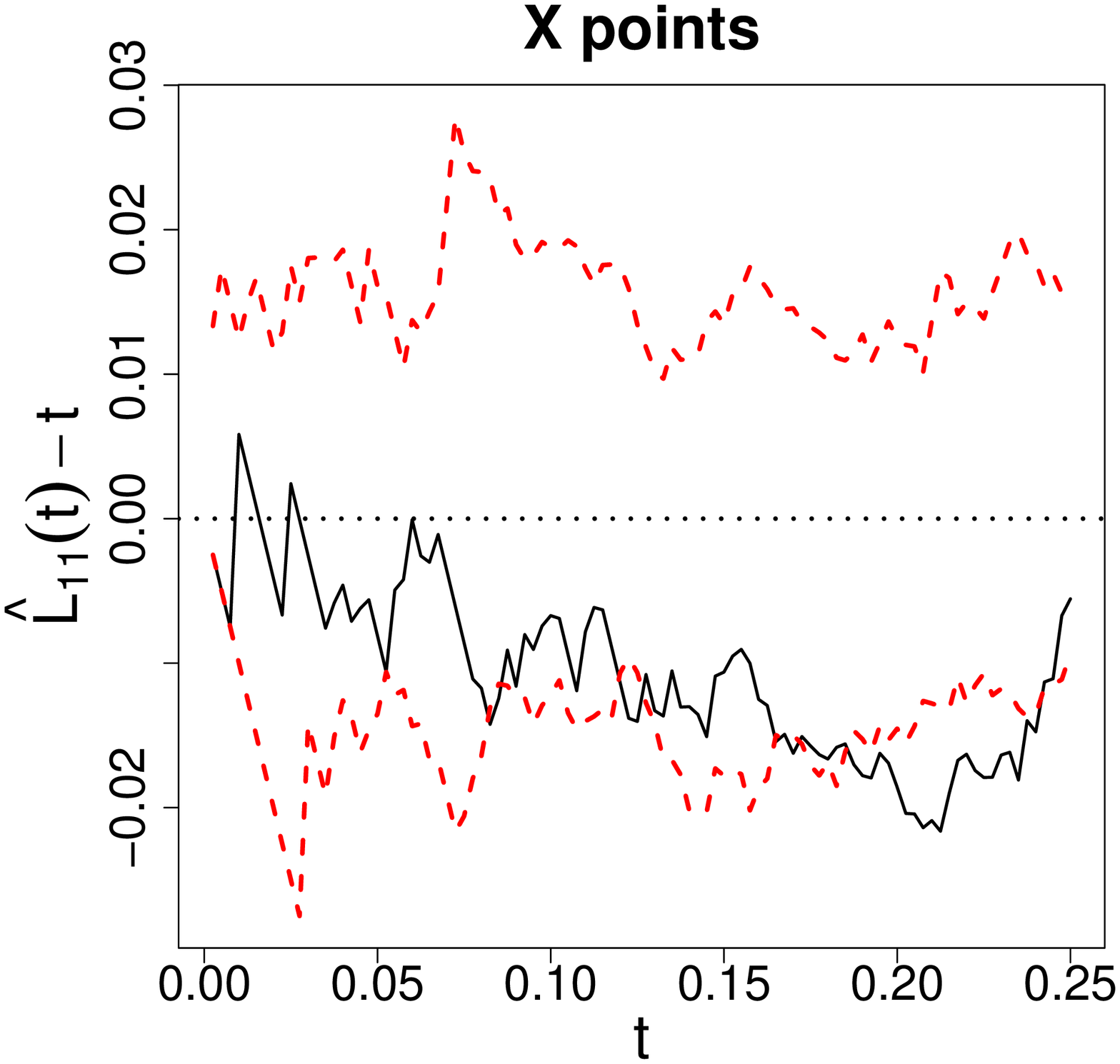} }}
\rotatebox{-90}{ \resizebox{2 in}{!}{\includegraphics{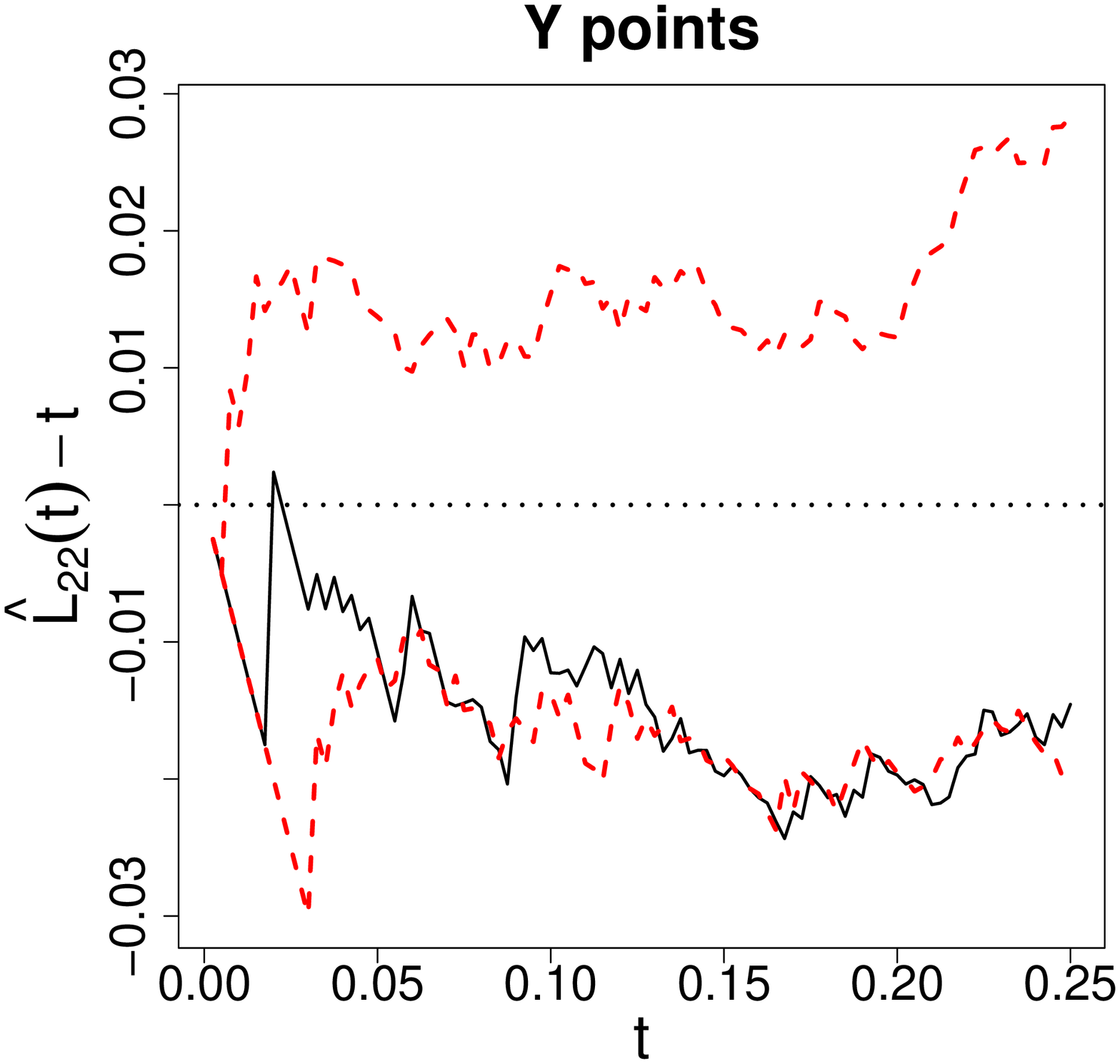} }}
\rotatebox{-90}{ \resizebox{2 in}{!}{\includegraphics{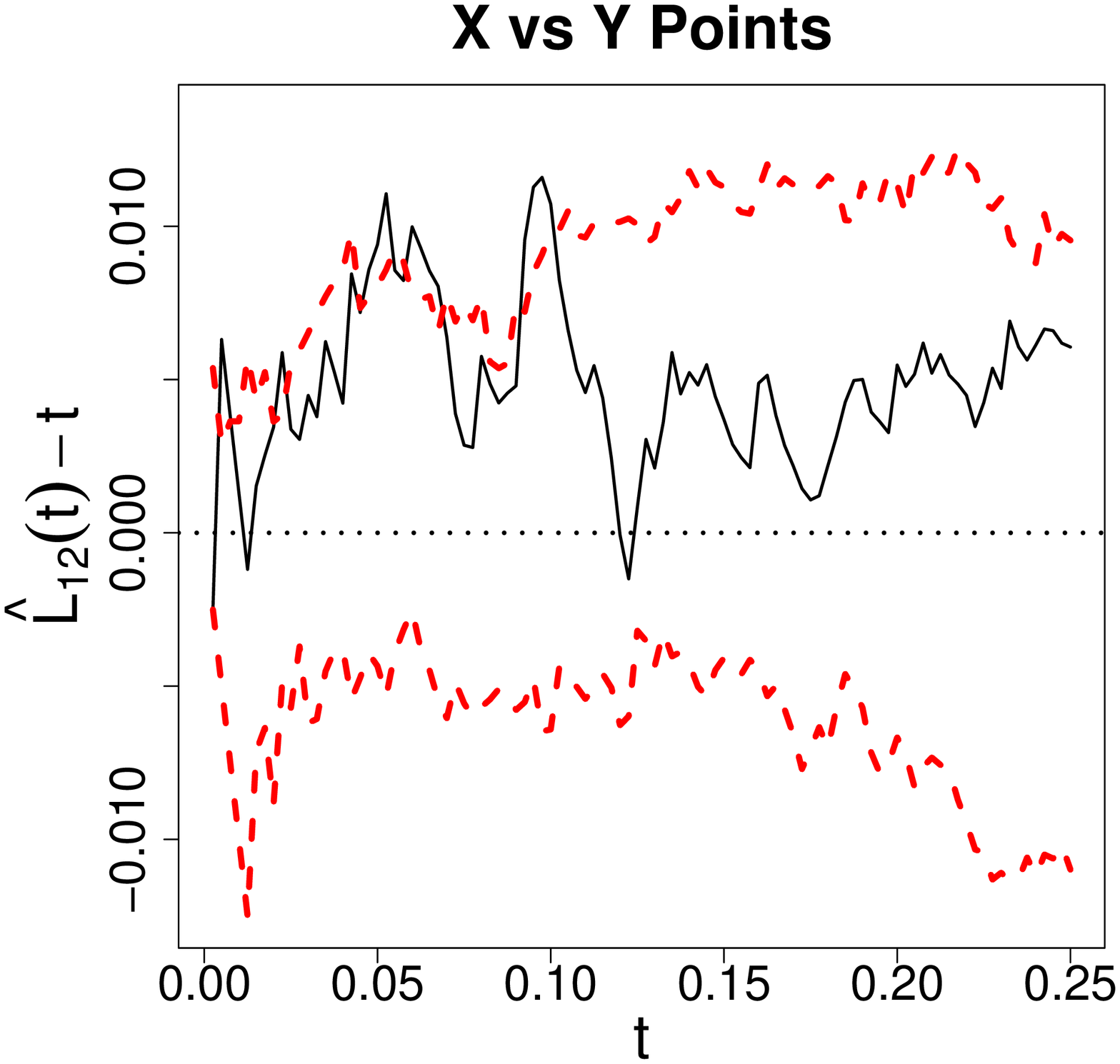} }}
\caption{
\label{fig:arti-Liihat}
Second-order properties of the artificial data.
Functions plotted are Ripley's univariate $L$-functions
$\widehat{L}_{ii}(t)-t$ for $i=1,2$,
and bivariate $L$-functions $\widehat{L}_{12}(t)-t$
where $i=1$ for $X$ and $i=2$ for $Y$ points.
The thick dashed lines around 0 are the upper and lower 95 \% confidence bounds for the
$L$-functions based on Monte Carlo simulation under the CSR independence pattern.
Notice also that vertical axes are differently scaled.}
\end{figure}

Since Pielou's and Dixon's tests give different results in terms of significance,
we also provide Ripley's univariate and bivariate $L$-functions
in Figure \ref{fig:arti-Liihat},
where the spatial interaction is analyzed for distances up to $t=0.25$.
Observe that $X$ points exhibit significant regularity
for distances $.18<t<.24$ and
no deviation from CSR for other distance values.
$Y$ points do not deviate significantly from CSR for all distances considered,
although they are close to being regular.
Observe also that
$X$ and $Y$ points are significantly associated for distances
around $t \approx 0.06$ and $t \approx 0.10$,
for all other distances their interaction does not
deviate significantly from CSR independence.
Hence we conclude that the significant segregation
implied by Pielou's test seems to be a false alarm,
since in fact, the spatial interaction between
the points is not significantly different from CSR independence.
Note also that average NN distance for the artificial data is
$.052 \pm .03$ and results of the
bivariate $L$-function and NNCT-analysis agree around $t=.05$.

\section{Discussion and Conclusions}
\label{sec:disc-conc}
In this article we discuss segregation tests based on
nearest neighbor contingency tables (NNCTs).
These NNCT-tests include Pielou's test (with or without Yates' correction),
Dixon's cell-specific and overall tests,
and the newly introduced one-sided versions of Pielou's tests.
A summary of the test statistics together with the underlying assumptions,
the appropriate null hypotheses, and the quantities the statistics are conditional on
are presented in Table \ref{tab:summary-test-stat}.

Pielou's and Dixon's tests were both defined under
the null hypothesis of random labeling (RL) of a fixed set of points
(\cite{pielou:1961},\cite{dixon:1994} and \cite{dixon:NNCTEco2002}).
It has been shown that Pielou's test is not appropriate for testing RL,
but Dixon's tests are.
The main problem with Pielou's NNCT-tests (including the one-sided versions)
discussed in this article is the dependence between trials
(i.e., categorization of (base, NN) pairs) and between the NNCT cell counts.

In this article, we extend the use of the NNCT-tests for the CSR of points
from two classes (i.e., CSR independence).
We demonstrate that Pielou's tests are liberal for rejecting RL or CSR independence,
while Dixon's tests are appropriate for large samples.
For smaller samples (i.e., when some cell count in the NNCT is $\le 5$)
we recommend the Monte Carlo randomization version for NNCT-tests.
We also show that Pielou's tests are only appropriate for a NNCT based
on a random sample of (base, nearest neighbor (NN)) pairs (which is not
a realistic situation in practice).
We prove the consistency of the tests under appropriate null hypotheses;
evaluate the empirical size performance of these tests based on an extensive
Monte Carlo simulation study under RL, CSR independence, and
independence of cell counts and rows in Section \ref{sec:Monte-Carlo}.
Based on the Monte Carlo analysis,
for moderate to large sample sizes Dixon's tests are recommended.

Under CSR independence, edge or boundary effects might be a potential
concern for the NNCT-tests.
Based on Monte Carlo analysis for edge correction methods,
with buffer zone correction,
we find that the uncorrected and corrected results are not significantly different.
In particular, buffer zone correction methods
(with larger distances than the average NN distances) are not recommended,
as they are wasteful procedures and do not serve the purpose of correcting for the edge effects.
The outer buffer zone correction with large buffer areas is redundant hence
not worth the effort,
while inner buffer zone correction with large buffer areas is wasteful
and might cause bias and loss of power in the analysis.
With toroidal correction
only significant change occurs for Dixon's overall change, but corrected versions
are usually liberal.
Hence edge effects are not significantly confounding the results
of the NNCT-tests (under CSR independence).

In this article, NNCTs are based on NN relations using the usual Euclidean distance.
But, this method can be extended to the case
that NN relation is based on dissimilarity measures
between observations in finite or infinite dimensional space.
It is even possible to have situations that are completely non-spatial
and yet one can conduct NNCT analysis based on dissimilarity measures.
In this general context the NN of object $x$ refers to the
object with the minimum dissimilarity to $x$.
The extension of RL pattern is straightforward,
but extra care should be taken for such an extension of CSR independence.
For example, in either case, the term $Q$ (in Dixon's tests) which is the number
of points with shared NNs needs to be revised
as $\widetilde Q:=2\,\sum_{k=1}^N {k \choose 2} Q_k$.
We conjecture that these tests when applied to other
fields for high dimensional data and NN relations
based on dissimilarity measures, can be useful.

{\small

\section*{Appendix}

\subsection*{Proof of Theorem \ref{thm:piel-directional} }
(I) Suppose under $H_o$, NNCT is based on a random sample of (base, NN) pairs.
For the two-class case,
we parametrize the segregation alternative 
as $H^S_a:\;\widetilde \pi_{11}=\widetilde \pi_{21}+\ve$ for $\ve \in (0,1-\widetilde \pi_{21})$.
Then the hypotheses become $H_o:\,\ve=0$ and $H^S_a:\;\ve>0$.
The rejection criterion in the theorem is equivalent to $Z_n>z(1-\al)$
where $Z_n$ is defined as in Equation \eqref{eqn:2x2-Zform}.
Then $\E[Z_n|H_o]=0$ and $\E\left[Z_n|H_a^S \right]=\ve>0$.

In the row-wise binomial framework with $N_{11} \sim
\BIN(n_1,\widetilde \pi_{11})$ and $N_{21} \sim \BIN(n_2,\widetilde \pi_{12})$ being
independent, consider $T_n=(N_{11}/n_1-N_{21}/n_2)$. 
Under $H_o$, $\E[T_n|H_o]=0$ and under $H^S_a$, the expected value
of $T_n$ becomes $\E\left[ T_n|H_a^S \right]=\ve>0$.
Under both null and
alternative hypotheses the test statistic
$(T_n-\E[T_n])/\sqrt{\Var[T_n]}$ has normal distribution
asymptotically.
By an appropriate application of Slutsky's Theorem,
$Z_n=(N_{11}/n_1-N_{21}/n_2)\,\sqrt{\frac{n_1\,n_2\,n}{C_1\,C_2}}$
and $(T_n-\E[T_n])/\sqrt{\Var[T_n]}$ have the same asymptotic
distribution.
Hence the size of the test is $\al$.
Furthermore, consistency follows by the asymptotic normality.
The consistency for
spatial association for the two-class case follows similarly.

In the multinomial framework with
$\mathbf{N}=(N_{11},N_{12},N_{21},N_{22})\sim \mathscr
M(n,\widetilde \nu_1\widetilde \kappa_1,\widetilde \nu_1\widetilde \kappa_2,
\widetilde \nu_2\widetilde \kappa_1,\widetilde \nu_2\widetilde \kappa_2)$, we
parametrize the segregation alternative as
$H^S_a:\,\E\left[N_{11}/n\right]=\widetilde \nu_1\,\widetilde \kappa_1 +\ve$.
Consider the test statistic $T_n:=N_{11}/n-\widetilde \nu_1\,\widetilde \kappa_1$.
Then expected value of $T_n$ under $H_o$ is $\E[T_n|H_o]=0$ and
$(T_n-\E[T_n])/\sqrt{\Var[T_n]}$ is approximately normal for large
$n$ under both null and alternative hypotheses.
By an appropriate application of Slutsky's Theorem and some algebraic manipulations,
one can see that $(T_n-\E[T_n])/\sqrt{\Var[T_n]}$ and
$\widetilde{Z}_n$ given in Equation \eqref{eqn:Zn-tilde} are
asymptotically equivalent and converge in law to the standard
normal distribution under $H_o$ and the same normal distribution under $H_a$.
So, the test has size $\al$ and using $T_n$,
consistency of the test for $H_a^S$ follows.
The consistency for spatial association for the two-class case follows similarly.

(II) Under RL or CSR independence,
$\displaystyle \E[T_n]=\frac{n_1-1}{n-1}-\frac{n_1}{n-1}=\frac{-1}{n-1}$ which
converges to zero as $n \rightarrow \infty$.
Let $\Var_{II}[T_n]$
be the variance of $T_n$ under RL or CSR independence and
$\Var_{I}[T_n]$ be the variance of $T_n$ under the case that (base, NN)
pairs are independent.
Then our claim (which is proved below) that
\begin{equation}
\label{eqn:var-compare}
\Var_{II}[T_n] > \Var_{I}[T_n]
\end{equation}
for large $n$.
Hence, $T_n$ rejects RL or CSR independence in
favor of segregation or spatial association more frequently than
it should; i.e., it is liberal in rejecting these null patterns,
or equivalently, its nominal significance level is larger than the
prespecified level $\al$.
However, under the above parametrization
of segregation, we have $\E\left[ T_n|H_a^S \right]=\ve>0$ and $\Var_{II}[T_n]$
converges to zero as $n \rightarrow \infty$.
Hence, consistency for segregation follows.
Consistency for the spatial association alternative follows similarly. $\blacksquare$

\subsubsection*{Proof of the Claim in Equation \eqref{eqn:var-compare}:}
For large $n$, the probabilities in Equation \eqref{eqn:probs} take
the form
$$p_{ii}=\nu_i^2,~~~ p_{ij}=\nu_i\,\nu_j, ~~~ p_{iii}= \nu_i^3,
~~~p_{iij}=\nu_i^2\,\nu_j,~~~ p_{iijj}=\nu_i^2\,\nu_j^2,~~~
p_{iiii}=\nu_i^4 \text{ for }i,j \in \{1,2\}.$$
Then variance of $T_n$ under RL or CSR independence is
$$\Var_{II}[T_n]=\Var[N_{11}/n_1]+\Var[N_{21}/n_2]-2\,\Cov[N_{11}/n_1,N_{21}/n_2].$$
Using Equation \eqref{eqn:VarNij} in Section
\ref{sec:dixon-cell-spec} and Equation (6) in
(\cite{dixon:NNCTEco2002}), we have,
\begin{multline*}
\Var[N_{11}/n_1]= \frac{1}{n^2\,\nu_1^2}
\left[(n+R)\,\nu_1^2+(2\,n-2\,R+Q)\,\nu_1^3+(n^2-3\,n-Q+R)\,\nu_1^4-(n\,\nu_1^2)^2\right] \\
=\frac{1}{n^2}
\left[n+R+(2\,n-2\,R+Q)\,\nu_1+(R-3\,n-Q)\,\nu_1^2\right]
\end{multline*}
and
\begin{multline*}
\Var[N_{21}/n_2]= \frac{1}{n^2\,\nu_2^2}
\left[n\,\nu_1\,\nu_2+Q\,\nu_2^2\,\nu_1+(n^2-3\,n-Q+R)\,\nu_1^2\,\nu_2^2
-(n\,\nu_1\,\nu_2)^2\right]\\
= \frac{1}{n^2\,\nu_2}
\left[n\,\nu_1+Q\,\nu_1\,\nu_2+(R-3\,n-Q)\,\nu_1^2\,\nu_2\right]
\end{multline*}
 and
\begin{multline*}
\Cov[N_{11}/n_1,N_{21}/n_2]=\frac{1}{n^2\,\nu_1\,\nu_2}
\left[(n-R-Q)\,\nu_1^2\,\nu_2+(n^2-3\,n-Q+R)\,\nu_1^3\,\nu_2-n^2\,\nu_1^3\,\nu_2\right]\\
=\frac{1}{n^2} \left[(n-R-Q)\,\nu_1+(R-3\,n-Q)\,\nu_1^2\right].
\end{multline*}
Hence by algebraic manipulations, we get
$$\Var_{II}[T_n]=\frac{n+R}{n^2}+\frac{\nu_1}{n\,\nu_2}.$$
Furthermore, under RL or CSR independence, $\kappa_i=\nu_i$ for $i=1,2$.
Then for large $n$, variance of $T_n$ under RL or CSR independence
is
\begin{align*}
\Var_{I}[T_n]&=\frac{\pi_{11}(1-\pi_{11})}{n_1}+\frac{\pi_{21}(1-\pi_{21})}{n_2}
=\frac{\nu_1\,\kappa_1(1-\nu_1\,\kappa_1)}{n\,\nu_1}+
\frac{\nu_2\,\kappa_1(1-\nu_2\,\kappa_1)}{n\,\nu_2}\\
&=\frac{\nu_1^2(1-\nu_1^2)}{n\,\nu_1}+
\frac{\nu_2\,\nu_1(1-\nu_2\,\nu_1)}{n\,\nu_2}
=\frac{\nu_1(1-\nu_1^2)}{n\,\nu_1}+
\frac{\nu_1(1-\nu_2\,\nu_1)}{n\,\nu_2}=\frac{\nu_1\,(2-\nu_1)}{n}.
\end{align*}
Then we need to show that,
$\displaystyle \frac{n+R}{n}+\frac{\nu_1}{\nu_2}=1+\frac{R}{n}+\frac{\nu_1}{\nu_2} > \nu_1\,(2-\nu_1)$
which trivially follows, since $1> \nu_1\,(2-\nu_1)$ which follows
from $(\nu_1-1)^2>0$. $\blacksquare$

\subsection*{Proof of Theorem \ref{thm:piel-chi2} }
(I) Suppose under $H_o$, NNCT is based on a random sample of (base, NN) pairs.
In the two-class case, deviations from $H_o$ are as in Theorem
\ref{thm:piel-directional}.
With such a deviation from $H_o$;
i.e., under $H_a$, for large $n$, $\X_P^2$ is approximately
distributed as a $\chi^2$ distribution with non-centrality
parameter $\lam(\ve)$ and $1$ degrees of freedom (d.f.), which is
denoted as $\chi^2_{1}(\lam(\ve))$.
The non-centrality parameter
is a quadratic form which can be written as
$\boldsymbol{\mu}(\ve)'\,A\,\boldsymbol{\mu}(\ve)$ for some
positive definite $2 \times 2$ matrix $A$ of rank $1$ (see
\cite{moser:1996}) hence $\lam(\mathbf{\ve})>0$ under $H_a$.
Then as $n \rightarrow \infty$,
the null and alternative hypotheses are
equivalent to $H_o:\,\lam=0$ versus $H_a:\,\lam=\lam(\ve)>0$.
Then the size of the test is $\al$ and the consistency follows.

\noindent
(II) Under RL or CSR independence, the dependence in the row sums
or column sums, which causes the reduction in d.f. is not the only
type of dependence present in the NNCT.
In addition to this, the
NNCT cell counts are also dependent due to reflexivity and shared NN structure.
Hence, even under RL or CSR independence, the
corresponding distribution is still a scaled version of central
$\chi^2$ distribution,
but has a larger variance than $\chi_1^2$
distribution, hence the nominal level of the test is larger than
the prespecified $\al$.
On the other hand, as $n \rightarrow
\infty$, the power of the test goes to 1. $\blacksquare$

\subsection*{Proof of Theorem \ref{thm:dixon-cell} }
 Consider the one-sided
alternative with $H_a:\,\pi_{ij}>\nu_{ij}$. Let
$T_n=N_{ij}/n-\nu_{ij}$. Then
$(T_n-\E[T_n])/\sqrt{\Var[T_n]}=Z^D_{ij}$.
Under RL,
$\E[T_n]=0$ and $\Var[T_n]$ is given in Equation \eqref{eqn:VarNij}.
Consider the parametrization of the
alternative as $H_a:\,\pi_{ij}=\nu_{ij}+\ve$ for $\ve \in
(0,1-\nu_{ij})$. Then under $H_a$, $\E[T_n|H_a]=\ve>0$.
As $n_i \rightarrow \infty$, $N_{ii}$ have asymptotic normal
distribution (\cite{cuzick:1990}).
This implies $N_{12}$ is also
asymptotically normal, as $n_1 \rightarrow \infty$, since
$N_{12}=n_1-N_{11}$.
Similarly, $N_{21}$ has asymptotically normal
distribution as $n_2 \rightarrow \infty$.
Thus under both null and
alternative hypothesis, $(T_n-\E[T_n])/\sqrt{\Var[T_n]}$ has
asymptotic normal distribution.
Then the size of the test is $\al$ and consistency follows.
The consistency for the other types of alternatives follow similarly.
$\blacksquare$

\subsection*{Proof of Theorem \ref{thm:dixon-overall} }
Under RL,
$\mathbf{Y}=\mathbf{N}-\E[\mathbf{N}]$ is approximately
distributed as $N(\mathbf{0},\Sigma)$ for large $n$.
Let $\Sigma^-$ be the generalized inverse of $\Sigma$ whose rank is $2$.
Then by Theorem 3.1.2 of \cite{moser:1996}, under $H_o$,
$\mathbf{Y}'\Sigma^{-}\mathbf{Y} \sim \chi^2_{2}(\lam=0)$.
Hence the test has size $\al$.
Consider any deviation from $H_o$.
Then under $H_a$, $\E[\mathbf{Y}|H_a]=\boldsymbol{\mu}_a$ and
$\mathbf{Y}$ have multivariate normal distribution with mean
$\boldsymbol{\mu}_a$.
Then by Theorem 3.1.2 of \cite{moser:1996},
under $H_a$, $\mathbf{Y}'\Sigma^{-}\mathbf{Y} \sim \chi^2_{2}
(\lam=\boldsymbol{\mu}_a'\Sigma^-\boldsymbol{\mu}_a)$.
Since $\Sigma^-$ is positive definite and $\boldsymbol{\mu}_a$ is
nonzero, the mean of the quadratic form $\X_D^2$ is $\lam+2$ with $\lam>0$.
So for large $N$, the null and alternative hypotheses
are equivalent to $H_o:\,\lam=0$ and $H_a:\,\lam>0$.
Then consistency follows. $\blacksquare$
}

\section*{Acknowledgments}
I would like to thank Prof. Philip M. Dixon and
an anonymous referee for their comments and
suggestions on earlier versions of this manuscript.
Most of the Monte Carlo simulations presented in this article
were executed on the Hattusas cluster of
Ko\c{c} University High Performance Computing Laboratory.



\section*{Biographical Sketch}
Elvan Ceyhan is an assistant professor in the Department of
Mathematics at Ko\c{c} University, Istanbul, Turkey.
He had a PhD from Applied
Mathematics and Statistics Department at Johns Hopkins University,
where he became interested in ecological statistics,
especially spatial point patterns.
In fact, his dissertation introduced a
graph theoretical method which has applications in spatial pattern analysis.

\section*{Tables and Figures}

\begin{table}[ht]
\centering
\begin{tabular}{cc|cc|c}
\multicolumn{2}{c}{}& \multicolumn{2}{c}{NN class}& \\
\multicolumn{2}{c}{}&    class 1 &  class $2$   &   total  \\
\hline
&class 1 &    $N_{11}$  &   $N_{12}$    &   $n_1$  \\
\raisebox{1.5ex}[0pt]{base class}
&class $2$ &    $N_{21}$ &  $N_{22}$    &   $n_2$  \\
\hline
&total     &    $C_1$   & $C_2$             &   n  \\
\end{tabular}
\caption{ \label{tab:NNCT-2x2} The NNCT for two classes.}
\end{table}

%

\begin{table}[ht]
\centering
\begin{tabular}{|c||c|c||c|c||c|c|c|}
\hline
\multicolumn{8}{|c|}{Empirical significance levels of the tests under} \\
\hline
\multicolumn{8}{|c|}{RL Case (1)} \\
\hline
 & \multicolumn{2}{|c||}{cell-specific} & \multicolumn{2}{|c||}{one-sided}
 & \multicolumn{3}{|c|}{overall} \\
\hline
$(n_1,n_2)$  & $\ah^D_{1,1}$ & $\ah^D_{2,2}$
  & $\ah_R$ & $\ah_L$ & $\ah_{P}$ & $\ah_{PY}$ & $\ah_{D}$ \\
\hline
$(182,91)$ & .0420$^c$ & .0487 & .1563$^\ell$ & .1874$^\ell$ & .1228$^\ell$ & .0972$^\ell$ & .0486 \\
\hline
\multicolumn{8}{|c|}{RL Case (2)} \\
\hline
$(10,10)$ & .0604$^\ell$ & .0557$^\ell$ & .0800$^\ell$ & .1479$^\ell$ & .1219$^\ell$ & .0586$^\ell$ & .0349$^c$ \\
\hline
$(10,30)$ & .0311$^c$ & .0699$^\ell$ & .0824$^\ell$ & .1456$^\ell$ & .1522$^\ell$ & .0618$^\ell$ & .0466 \\
\hline
$(10,50)$ & .0264$^c$ & .0472 & .0953$^\ell$ & .0517 & .0640$^\ell$ & .0307$^c$ & .0507 \\
\hline
$(30,30)$ & .0579$^\ell$ & .0547$^\ell$ & .0749$^\ell$ & .1065$^\ell$ & .1249$^\ell$ & .0805$^\ell$ & .0497 \\
\hline
$(30,50)$ & .0621$^\ell$ & .0608$^\ell$ & .0896$^\ell$ & .1203$^\ell$ & .1338$^\ell$ & .0826$^\ell$ & .0444$^c$ \\
\hline
$(50,50)$ & .0512 & .0524 & .0794$^\ell$ & .1058$^\ell$ & .1383$^\ell$ & .1025$^\ell$ & .0497 \\
\hline
$(50,100)$ & .0625$^\ell$ & .0512 & .0905$^\ell$ & .1060$^\ell$ & .1199$^\ell$ & .0926$^\ell$ & .0482 \\
\hline
$(100,100)$ & .0538$^\ell$ & .0534 & .0895$^\ell$ & .1101$^\ell$ & .1321$^\ell$ & .1052$^\ell$ & .0525 \\
\hline
\multicolumn{8}{|c|}{RL Case (3)} \\
\hline
$(10,10)$ & .0624$^\ell$ & .0657$^\ell$ & .0806$^\ell$ & .1492$^\ell$ & .1220$^\ell$ & .0571$^\ell$ & .0446$^c$ \\
\hline
$(10,30)$ & .0297$^c$ & .0341$^c$ & .0803$^\ell$ & .1454$^\ell$ & .1382$^\ell$ & .0517 & .0327$^c$ \\
\hline
$(10,50)$ & .0251$^c$ & .0384$^c$ & .0882$^\ell$ & .0463$^c$ & .0591$^\ell$ & .0287$^c$ & .0508 \\
\hline
$(30,30)$ & .0513 & .0523 & .0839$^\ell$ & .1179$^\ell$ & .1402$^\ell$ & .0933$^\ell$ & .0469 \\
\hline
$(30,50)$ & .0626$^\ell$ & .0594$^\ell$ & .0934$^\ell$ & .1174$^\ell$ & .1367$^\ell$ & .0846$^\ell$ & .0411$^c$ \\
\hline
$(50,50)$ & .0509 & .0511 & .0800$^\ell$ & .1113$^\ell$ & .1414$^\ell$ & .1047$^\ell$ & .0501 \\
\hline
$(50,100)$ & .0566$^\ell$ & .0421$^c$ & .0906$^\ell$ & .1019$^\ell$ & .1182$^\ell$ & .0906$^\ell$ & .0460$^c$ \\
\hline
$(100,100)$ & .0439$^c$ & .0453$^c$ & .0942$^\ell$ & .1127$^\ell$ & .1361$^\ell$ & .1098$^\ell$ & .0505 \\
\hline
\multicolumn{8}{|c|}{RL Case (4)} \\
\hline
$(10,10)$ & .0656$^\ell$ & .0640$^\ell$ & .0798$^\ell$ & .1481$^\ell$ & .1236$^\ell$ & .0536 & .0432$^c$ \\
\hline
$(10,30)$ & .0281$^c$ & .0447$^c$ & .0798$^\ell$ & .1525$^\ell$ & .1639$^\ell$ & .0521 & .0324$^c$ \\
\hline
$(10,50)$ & .0260$^c$ & .0404$^c$ & .0892$^\ell$ & .0506 & .0618$^\ell$ & .0290$^c$ & .0500 \\
\hline
$(30,30)$ & .0549$^\ell$ & .0553$^\ell$ & .0858$^\ell$ & .1183$^\ell$ & .1459$^\ell$ & .0984$^\ell$ & .0484 \\
\hline
$(30,50)$ & .0677$^\ell$ & .0685$^\ell$ & .0936$^\ell$ & .1156$^\ell$ & .1359$^\ell$ & .0861$^\ell$ & .0445$^c$ \\
\hline
$(50,50)$ & .0504 & .0506 & .0769$^\ell$ & .1094$^\ell$ & .1372$^\ell$ & .0991$^\ell$ & .0488 \\
\hline
$(50,100)$ & .0590$^\ell$ & .0484 & .0882$^\ell$ & .1006$^\ell$ & .1179$^\ell$ & .0887$^\ell$ & .0479 \\
\hline
$(100,100)$ & .0495 & .0476 & .0941$^\ell$ & .1134$^\ell$ & .1406$^\ell$ & .1137$^\ell$ & .0534 \\
\hline
\end{tabular}
\caption{ \label{tab:overall-null-RL}
The empirical significance levels
for the under RL cases (1)-(4) at $\alpha=.05$.
Here $\ah_{i,i}^D$ is the empirical significance level for
Dixon's cell-specific test for cell $(i,i)$ for $i \in \{1,2\}$,
$\ah_R$ is for the right-sided version of Pielou's test,
$\ah_L$ is for left-sided version of Pielou's test,
$\ah_{P}$ and $\ah_{PY}$ are for Pielou's overall test of
segregation without and with Yates' correction, respectively,
$\ah_{D}$ is for Dixon's overall test of segregation.
$^c$:The empirical size is significantly smaller than .05;
i.e., the test is conservative.
$^\ell$:The empirical size is significantly larger than .05;
i.e., the test is liberal.}
\end{table}

\begin{table}[ht]
\centering
\begin{tabular}{|c||c|c|c|c||c|c||c|c|c|c|}
\hline
\multicolumn{11}{|c|}{Empirical significance levels under CSR independence} \\
\hline
&\multicolumn{4}{|c||}{cell-specific} & \multicolumn{2}{|c||}{one-sided}
& \multicolumn{4}{|c|}{overall} \\
\hline
$(n_1,n_2)$  & $\ah_{1,1}^D$ & $\ah_{1,1}^{D,qr}$ &  $\ah_{2,2}^D$ &  $\ah_{2,2}^{D,qr}$ & $\ah_R$ & $\ah_L$ & $\ah_{P}$ &
$\ah_{PY}$ & $\ah_{D}$ & $\ah_{D,qr}$ \\
\hline
$(10,10)$ & .0454$^c$ & .0360$^c$ & .0465 & .0383$^c$ & .0844$^\ell$ & .1574$^\ell$ & .1280$^\ell$ & .0608$^\ell$ & .0432$^c$ & .0470 \\
\hline
$(10,30)$ & .0306$^c$ & .0306$^c$ & .0485 & .0427$^c$  & .0846$^\ell$ &  .1399$^\ell$ & .1429$^\ell$ & .0542$^\ell$ & .0440$^c$ & .0411$^c$ \\
\hline
$(10,50)$ & .0270$^c$ & .0270$^c$ & .0464 & .0323$^c$  & .0947$^\ell$ &  .0574$^\ell$ &  .0664$^\ell$ & .0318$^c$ & .0482 & .0497  \\
\hline
$(30,10)$ & .0479 & .0415$^c$ & .0275$^c$ & .0275$^c$  & .0760$^\ell$ & .1406$^\ell$ & .1383$^\ell$ &.0506 & .0390$^c$ & .0402$^c$  \\
\hline
$(30,30)$ & .0507 & .0577$^\ell$ & .0505 & .0578$^\ell$ & .0803$^\ell$ & .1115$^\ell$ & .1339$^\ell$ & .0836$^\ell$ & .0464 & .0492  \\
\hline
$(30,50)$ & .0590$^\ell$ & .0591$^\ell$ & .0522 & .0549$^\ell$ & .0821$^\ell$ & .1211$^\ell$ & .1319$^\ell$ & .0834$^\ell$ & .0454$^c$ & .0411$^c$  \\
\hline
$(50,10)$ & .0524 & .0346$^c$ & .0263$^c$ & .0263$^c$ & .0955$^\ell$ &  .0544$^\ell$ &  .0654$^\ell$ & .0310$^c$ & .0529 & .0510  \\
\hline
$(50,30)$ & .0535 & .0554$^\ell$ & .0597$^\ell$ & .0597$^\ell$ & .0829$^\ell$ & .1173$^\ell$ & .1275$^\ell$ & .0805$^\ell$ & .0429$^c$ & .0405$^c$ \\
\hline
$(50,50)$ & .0465 & .0456$^c$ & .0469 & .0459$^c$ & .0804$^\ell$ & .1041$^\ell$ & .1397$^\ell$ & .0999$^\ell$ & .0508 & .0528  \\
\hline
$(50,100)$ & .0601$^\ell$ & .0652$^\ell$ & .0533 & .0535 & .0921$^\ell$ & .1090$^\ell$ & .1223$^\ell$ & .0938$^\ell$ & .0560$^\ell$ & .0556$^\ell$  \\
\hline
$(100,50)$ & .0490 & .0493 & .0571$^\ell$ & .0620$^\ell$ & .0909$^\ell$ & .1063$^\ell$ & .1190$^\ell$ & .0904$^\ell$ & .0483 & .0495  \\
\hline
$(100,100)$ & .0493 & .0491 & .0463$^c$ & .0455 & .0927$^\ell$ & .1092$^\ell$ & .1324$^\ell$ & .1076$^\ell$ & .0504 & .0513  \\
\hline
\end{tabular}
\caption{ \label{tab:null-CSR}
The empirical significance levels
for the tests under $H_o:$ CSR independence with $N_{mc}=10000$, $n_1,n_2 \in
\{10,30,50,100\}$ at $\alpha=.05$ for uniform class $X$ and $Y$
points in the unit square.
Here $\ah_{i,i}^{D,qr}$ is the empirical significance level for
the QR-adjusted cell-specific test for $(i,i)$ for $i \in \{1,2\}$,
and $\ah_{D,qr}$ is for QR-adjusted Dixon's overall test of segregation.
Other empirical size estimate notation and superscript labeling are as in
Table \ref{tab:overall-null-RL}.
}
\end{table}

\begin{table}[ht]
\centering
\begin{tabular}{|c||c|c|c|c|c|c|c|c|}
\hline
\multicolumn{9}{|c|}{Proportion of agreement between Pielou's and Dixon's overall tests} \\
\hline
$(n_1,n_2)$  & $(10,10)$ & $(10,30)$ &  $(10,50)$ &  $(30,30)$ & $(30,50)$ & $(50,50)$ &
$(50,100)$& $(100,100)$ \\
\hline
RL Case (2) & .0255 & .0245 & .0344 & .0317 & .0275 & .0321 & .0302 & .0336  \\
\hline
RL Case (3) & .0226 & .0189 & .0321 & .0323  & .0275 &  .0331 & .0289 & .0320  \\
\hline
RL Case (4) & .0208 & .0192 & .0335 & .0329  & .0250 &  .0292 &  .0298 & .0355 \\
\hline
CSR independence & .0277 & .0242 & .0340 & .0296  & .0270 & .0321 & .0342 &.0314   \\
\hline
\end{tabular}
\caption{ \label{tab:prop-agree}
The proportion of agreement between Pielou's and Dixon's overall segregation tests
$\ah_{P,D}$ for rejecting the RL Cases (2)-4 and CSR independence with $N_{mc}=10000$, $n_1,n_2 \in
\{10,30,50,100\}$ at $\alpha=.05$.
}
\end{table}

\begin{table}[ht]
\centering
\begin{tabular}{|c||c|c||c|c||c|c||c|c|}
\hline
\multicolumn{9}{|c|}{Empirical significance levels under independence of cells} \\
\hline
& \multicolumn{4}{|c||}{overall multinomial} & \multicolumn{4}{c|}{row-wise binomial} \\
\hline
& \multicolumn{2}{|c||}{one-sided} & \multicolumn{2}{|c||}{overall}
& \multicolumn{2}{|c||}{one-sided} & \multicolumn{2}{c|}{overall} \\
\hline
$(n_1,n_2)$  & $\ah_R$ & $\ah_L$ & $\ah_{P}$ & $\ah_{PY}$  &
$\ah_R$ & $\ah_L$ & $\ah_{P}$ & $\ah_{PY}$ \\
\hline
$(10,10)$ & .0542$^\ell$ & .0521 & .0415$^c$ & .0102$^c$ & .0603$^\ell$ & .0612$^\ell$ & .0426$^c$ & .0124$^c$ \\
\hline
$(10,30)$ & .0535 & .0560$^\ell$ & .0525 & .0213$^c$ & .0575$^\ell$ & .0495 & .0510 & .0174$^c$ \\
\hline
$(10,50)$ & .0489 & .0534 & .0556$^\ell$ & .0261$^c$ & .0864$^\ell$ & .0000$^c$ & .0468 & .0152$^c$ \\
\hline
$(30,10)$ & .0559$^\ell$ & .0511 & .0500 & .0187$^c$ & .0575$^\ell$ & .0453$^c$ & .0504 & .0186$^c$ \\
\hline
$(30,30)$ & .0470 & .0445$^c$ & .0500 & .0268$^c$ & .0487 & .0491 & .0518 & .0269$^c$ \\
\hline
$(30,50)$ & .0493 & .0504 & .0492 & .0272$^c$ & .0537$^\ell$ & .0541 & .0554$^\ell$ & .0263$^c$ \\
\hline
$(50,10)$ & .0579$^\ell$ & .0440$^c$ & .0551$^\ell$ & .0254$^c$ & .0707$^\ell$ & .0205$^c$ & .0410$^c$ & .0132$^c$ \\
\hline
$(50,30)$ & .0504 & .0493 & .0492 & .0272$^c$ & .0539$^\ell$ & .0531 & .0526 & .0274$^c$ \\
\hline
$(50,50)$ & .0483 & .0487 & .0534 & .0344$^c$ & .0451$^c$ & .0466 & .0592$^\ell$ & .0357$^c$ \\
\hline
$(50,100)$ & .0495 & .0493 & .0503 & .0336$^c$ & .0509 & .0487 & .0488 & .0333$^c$ \\
\hline
$(100,50)$ & .0493 & .0495 & .0503 & .0336$^c$ & .0512 & .0460$^c$ & .0510 & .0350$^c$ \\
\hline
$(100,100)$ & .0499 & .0500 & .0537$^c$ & .0379$^c$ & .0528 & .0490 & .0538$^\ell$ & .0378$^c$\\
\hline
\end{tabular}
\caption{ \label{tab:null-indep}
The empirical significance levels
for the tests under independence of cells and rows with $N_{mc}=10000$, $n_1,n_2 \in
\{10,30,50,100\}$ at $\alpha=.05$ for contingency tables
based on the overall multinomial and row-wise binomial frameworks.
The empirical size notation and superscript labeling are as in
Table \ref{tab:overall-null-RL}.}
\end{table}

\begin{table}[ht]
\centering
\begin{tabular}{|c||c|c||c|c||c|c|c|}
\hline
\multicolumn{8}{|c|}{Empirical significance levels for the NNCT-tests with buffer zone correction} \\
\hline
 & \multicolumn{2}{|c||}{cell-specific} & \multicolumn{2}{|c||}{one-sided}
 & \multicolumn{3}{|c|}{overall} \\
\hline
$(n_1,n_2)$  & $\ah_{1,1}^D$  & $\ah_{2,2}^D$ & $\ah_R$ & $\ah_L$ & $\ah_{P}$ & $\ah_{PY}$ & $\ah_{D}$ \\
\hline
$(10,10)$ & .0464$^{\approx}$ & .0442$^{c,\approx}$ & .0811$^{\ell,\approx}$ &
.1402$^{\ell,<}$ & .1169$^{\ell,<}$ & .0506$^<$ & .0604$^{\ell,>}$ \\
\hline
$(10,30)$ & .0317$^{c,\approx}$ & .0602$^{\ell,>}$ & .0830$^{\ell,\approx}$ &
.1296$^{\ell,<}$ & .1283$^{\ell,<}$ & .0530$^\approx$ & .0572$^{\ell,>}$ \\
\hline
$(10,50)$ & .0258$^{c,\approx}$ & .0526$^{>}$ & .0946$^{\ell,\approx}$ &
.0502$^<$ &  .0625$^{\ell,\approx}$ & .0294$^{c,\approx}$ & .0569$^{\ell,>}$ \\
\hline
$(30,10)$ & .0565$^{\ell,>}$ & .0289$^{c,\approx}$ & .0799 $^{\ell,\approx}$ &
.1273$^{\ell,<}$ & .1235$^{\ell,<}$ & .0485$^{\approx}$ & .0540$^{\ell,>}$ \\
\hline
$(30,30)$ & .0518$^{\approx}$ & .0522$^{\approx}$ & .0782$^{\ell,\approx}$ &
.1104$^{\ell,\approx}$ & .1284$^{\ell,\approx}$ & .0848$^{\ell,\approx}$ & .0545$^{\ell,>}$ \\
\hline
$(30,50)$ & .0608$^{\ell,\approx}$ & .0554$^{\ell,\approx}$ & .0886$^{\ell,\approx}$ &
.1088$^{\ell,<}$ & .1253$^{\ell,\approx}$ & .0769$^{\ell,\approx}$ & .0521$^>$ \\
\hline
$(50,10)$ & .0537$^{\ell,\approx}$ & .0269$^{c,\approx}$ & .0959$^{\ell,\approx}$ &
.0553$^{\ell,\approx}$ & .0669$^{\ell,\approx}$ & .0297$^{\ell,\approx}$ & .0599$^{\ell,>}$ \\
\hline
$(50,30)$ & .0573$^{\ell,\approx}$ & .0640$^{\ell,\approx}$ & .0870$^{\ell,\approx}$ &
.1122$^{\ell,\approx}$ & .1257$^{\ell,\approx}$ & .0822$^{\ell,\approx}$ & .0540$^{\ell,>}$ \\
\hline
$(50,50)$ & .0463$^{c,\approx}$ & .0471$^{\approx}$ & .0820$^{\ell,\approx}$ &
.0998$^{\ell,\approx}$ & .1330$^{\ell,\approx}$ & .0946$^{\ell,\approx}$ & .0510$^{\approx}$ \\
\hline
$(50,100)$ & .0513$^{<}$ & .0519$^{\approx}$ & .0939$^{\ell,\approx}$ &
.1040$^{\ell,\approx}$ & .1209$^{\ell,\approx}$ & .0905$^{\ell,\approx}$ & .0544$^{\ell,\approx}$ \\
\hline
$(100,50)$ & .0527$^{\approx}$ & .0497$^{<}$ & .0906$^{\ell,\approx}$ &
.1038$^{\ell,\approx}$ & .1211$^{\ell,\approx}$ & .0903$^{\ell,\approx}$ & .0481$^{\approx}$ \\
\hline
$(100,100)$ & .0473$^{\approx}$ & .0475$^{\approx}$ & .0922$^{\ell,\approx}$ &
.1069$^{\ell,\approx}$ & .1314$^{\ell,\approx}$ & .1039$^{\ell,\approx}$ & .0493$^{\approx}$ \\
\hline
\end{tabular}
\caption{ \label{tab:overall-buff} The empirical significance levels
for the overall tests under $H_o$ with $N_{mc}=10000$, $n_1,n_2 \in
\{10,20,30,40,50\}$ at $\alpha=.05$ for uniform class $X$ and $Y$ points
in the unit square when edge correction with buffer zone is applied.
The empirical size notation is as in
Table \ref{tab:overall-null-RL}.
$^c$:The empirical size is significantly smaller than .05; i.e., the test is conservative.
$^\ell$:The empirical size is significantly larger than .05; i.e., the test is liberal.
$^<$($^>$):Empirical size significantly smaller (larger) than the uncorrected size.
$^\approx$: Empirical size not significantly different from the uncorrected size.}
\end{table}

\begin{table}[ht]
\centering
\begin{tabular}{|c||c|c||c|c||c|c|c|}
\hline \multicolumn{8}{|c|}{Empirical significance levels for the NNCT-tests with toroidal edge correction} \\
\hline
 & \multicolumn{2}{|c||}{cell-specific} & \multicolumn{2}{|c||}{one-sided}
 & \multicolumn{3}{|c|}{overall} \\
\hline
$(n_1,n_2)$  & $\ah_{1,1}^D$  & $\ah_{2,2}^D$ & $\ah_R$ & $\ah_L$ & $\ah_{P}$ & $\ah_{PY}$ & $\ah_{D}$ \\
\hline
$(10,10)$ & .0414$^{c,\approx}$ & .0430$^{c,\approx}$ & .0782$^{\ell,\approx}$ &
.1531$^{\ell,\approx}$ & .1285$^{\ell,\approx}$ & .0620$^{\ell,\approx}$ & .0413$^{c,<}$ \\
\hline
$(10,30)$ & .0318$^{c,\approx}$ & .0492$^{\approx}$ & .0845$^{\ell,\approx}$ &
.1397$^{\ell,\approx}$ & .1434$^{\ell,\approx}$ & .0536$^\approx$ & .0383$^{c,\approx}$ \\
\hline
$(10,50)$ & .0265$^{c,\approx}$ & .0466$^{\approx}$ & .0958$^{\ell,\approx}$ &
.0561$^{\ell,\approx}$ &  .0670$^{\ell,\approx}$ & .0323$^{c,\approx}$ & .0490$^\approx$ \\
\hline
$(30,10)$  & .0453$^{c,\approx}$ & .0285$^{c,\approx}$ & .0777$^{\ell,\approx}$ &
.1412$^{\ell,\approx}$ & .1412$^{\ell,\approx}$ & .0508$^{\approx}$ & .0376$^{c,\approx}$ \\
\hline
$(30,30)$ & .0494$^{\approx}$ & .0474$^{\approx}$ & .0800$^{\ell,\approx}$ &
.1147$^{\ell,\approx}$ & .1338$^{\ell,\approx}$ & .0871$^{\ell,\approx}$ & .0447$^{c,\approx}$ \\
\hline
$(30,50)$ & .0592$^{\ell,\approx}$ & .0500$^{\approx}$ & .0832$^{\ell,\approx}$ &
.1127$^{\ell,\approx}$ & .1233$^{\ell,\approx}$ & .0783$^{\ell,\approx}$ & .0437$^{c,\approx}$ \\
\hline
$(50,10)$ & .0481$^{\approx}$ & .0267$^{c,\approx}$ & .0969$^{\ell,\approx}$ &
.0549$^{\ell,\approx}$ & .0660$^{\ell,\approx}$ & .0309$^{c,\approx}$ & .0499$^\approx$ \\
\hline
$(50,30)$ & .0518$^{\approx}$ & .0604$^{\ell,\approx}$ & .0827$^{\ell,\approx}$ &
.1204$^{\ell,\approx}$ & .1297$^{\ell,\approx}$ & .0785$^{\ell,\approx}$ & .0447$^{c,\approx}$ \\
\hline
$(50,50)$ & .0457$^{c,\approx}$  & .0444$^{c,\approx}$  & .0804$^{\ell,\approx}$ & .1026$^{\ell,\approx}$ &
.1401$^{\ell,\approx}$ & .1009$^{\ell,\approx}$ & .0454$^{c,\approx}$ \\
\hline
$(50,100)$ & .0533$^{<}$ & .0518$^{\approx}$ & .0923$^{\ell,\approx}$ & .1114$^{\ell,\approx}$ &
.1260$^{\ell,\approx}$ & .0958$^{\ell,\approx}$ & .0528$^\approx$ \\
\hline
$(100,50)$ & .0503$^{\approx}$ & .0522$^{\approx}$ & .0903$^{\ell,\approx}$ & .1079$^{\ell,\approx}$ &
.1245$^{\ell,\approx}$ & .0946$^{\ell,<}$ & .0485$^\approx$ \\
\hline
$(100,100)$ & .0487$^{\approx}$ & .0465$^{\approx}$ & .0954$^{\ell,\approx}$ & .1069$^{\ell,\approx}$ &
.1344$^{\ell,\approx}$ & .1077$^{\ell,<}$ & .0467$^\approx$ \\
\hline
\end{tabular}
\caption{ \label{tab:overall-tor}
The empirical significance levels
for the overall tests under $H_o$ with $N_{mc}=10000$, $n_1,n_2 \in
\{10,20,30,40,50\}$ at $\alpha=.05$ for uniform class $X$ and $Y$ points
in the unit square when toroidal edge correction is applied.
The empirical size notation and superscript labeling are as in
Table \ref{tab:overall-buff}.}
\end{table}

\begin{table}[ht]
\centering
%
%
Swamp Tree Data\\
\begin{tabular}{cc|cc|c}
\multicolumn{2}{c}{}& \multicolumn{2}{c}{NN}& \\
\multicolumn{2}{c}{}&    B.G. &  B.C.   &   sum  \\
\hline
& B.G.&    149  &   33    &   182  \\
\raisebox{1.5ex}[0pt]{base}
& B.C. &    43 &  48    &   91  \\
\hline
&sum     &    192   & 581            &  273  \\
\end{tabular}
\hspace{.1in}
\begin{tabular}{cc|cc|c}
\multicolumn{2}{c}{}& \multicolumn{2}{c}{NN}& \\
\multicolumn{2}{c}{}&    B.G. &  B.C.   &   sum  \\
\hline
&B.G.&   82 \%  &   18 \%    &   67 \%  \\
\raisebox{1.5ex}[0pt]{base}
&B.C. &    47 \% &  53 \%    &   23 \%  \\
\hline
&sum     &    34 \%   & 66 \%             &  100 \%  \\
\end{tabular}

\vspace{.2in}
Artificial Data\\
\begin{tabular}{cc|cc|c}
\multicolumn{2}{c}{}& \multicolumn{2}{c}{NN}& \\
\multicolumn{2}{c}{}&    $X$ &  $Y$   &   sum  \\
\hline
& $X$ &    30  &   20    &   50  \\
\raisebox{1.5ex}[0pt]{base}
& $Y$ &    19 &  31    &   50  \\
\hline
&sum     &    49   & 51            &  100  \\
\end{tabular}
\hspace{.1in}
\begin{tabular}{cc|cc|c}
\multicolumn{2}{c}{}& \multicolumn{2}{c}{NN}& \\
\multicolumn{2}{c}{}&    $X$ &  $Y$   &   sum  \\
\hline
& $X$ &   \% 60  &   \% 40    &   \% 50  \\
\raisebox{1.5ex}[0pt]{base}
& $Y$ &    \% 38 &  \% 62    &   \% 50  \\
\hline
&sum     &    \% 49   & \% 51             &  \% 100  \\
\end{tabular}
\caption{ \label{tab:NNCT-example}
The NNCT for swamp tree data (top left)
and the artificial data (bottom left) and the corresponding percentages (right).
B.G. = black gums, B.C. = bald cypresses.}
\end{table}

\begin{table}[ht]
\centering
\begin{tabular}{|c||c|c||c||c|c|c|}
\hline
%
\multicolumn{7}{|c|}{Test statistics and the associated $p$-values for swamp tree data} \\
\hline
 & $Z^D_{11}$ & $Z^D_{22}$  & $Z_n$ & $\X^2_P$ & $\X^2_{PY}$ & $\X^2_D$ \\
\hline
correction & \multicolumn{2}{|c||}{cell-specific} & \multicolumn{1}{|c||}{one-sided}
 & \multicolumn{3}{|c|}{overall} \\
\hline
none & 4.47 & 3.54 & 5.90 & 34.84 & 33.20  & 23.77 \\
     & $(<.0001)$ & $(.0004)$ & $(\approx 1,\,<.0001)$ & $(<.0001)$ & $(<.0001)$  & $(<.0001)$ \\
\hline
toroidal & 4.31 & 3.31 & 5.62 & 31.60 & 30.04 & 21.29 \\
     & $(<.0001)$ & $(.0009)$ & $(\approx 1,\,<.0001)$ & $(<.0001)$ & $(<.0001)$  & $(<.0001)$ \\
\hline
buffer zone & 3.95 & 4.04 & 6.08 & 37.00 & 35.10 & 23.39 \\
 ($k=0$)    & $(.0001)$ & $(.0001)$ & $(\approx 1,\,<.0001)$ & $(<.0001)$ & $(<.0001)$  & $(<.0001)$ \\
\hline
buffer zone & 3.61 & 4.08 & 5.90 & 34.77 & 32.86 & 21.92 \\
 ($k=1$)    & $(.0003)$ & $(.0001)$ & $(\approx 1,\,<.0001)$ & $(<.0001)$ & $(<.0001)$  & $(<.0001)$ \\
\hline
\multicolumn{7}{|c|}{Test statistics and the associated $p$-values for the artificial data} \\
\hline
none & 1.38 & 1.64 & 2.20 & 4.84 & 4.00  & 3.36 \\
     & $(.1670)$ & $(.1000)$ & $(.9861,\,.0139)$ & $(.0278)$ & $(.0455)$  & $(.1868)$ \\
\hline
\hline
toroidal & 1.38 & 1.38 & 2.00 & 4.00 & 3.24 & 2.65 \\
     & $(.1672)$ & $(.1672)$ & $(.9772,\,.0228)$ & $(.0455)$ & $(.0719)$  & $(.2660)$ \\
\hline
buffer zone & 1.07 & 1.28 & 1.72 & 2.95 & 2.22 & 2.09 \\
 ($k=0$)    & $(.2843)$ & $(.2004)$ & $(.9572,\,.0428)$ & $(.0857)$ & $(.1365)$  & $(.3511)$ \\
\hline
buffer zone & 0.61 & 0.58 & 0.81 & 0.66 & 0.32 & 0.54 \\
 ($k=1$)    & $(.5391)$ & $(.5597)$ & $(.7922,\,.2078)$ & $(.4156)$ & $(.5697)$  & $(.7637)$ \\
\hline
\end{tabular}
\caption{ \label{tab:example-test-stat}
The values of the NNCT-test statistics and the corresponding $p$-values (in parenthesis) for
the swamp tree data (top) and artificial data set (bottom).
$\X^2_D$: Dixon's overall segregation test,
$\X^2_P$ and $\X^2_{PY}$: Pielou's test without and
with Yates' correction, respectively,
$Z_n$: directional $Z$-test.
The $p$-values are for the general alternative of any deviation from CSR independence
except for $Z_n$, for which the first $p$-value in the parenthesis
is for the association alternative,
while the second is for the segregation alternative.
}
\end{table}

\begin{sidewaystable}
\centering
\begin{tabular}{|c|c|c|c|c|c|}
\hline
\multicolumn{6}{|c|}{A Summary of the NNCT-Tests} \\
\hline
test & underlying & appropriate & conditional & \multicolumn{2}{|c|}{empirical size under}\\
statistic & assumption & $H_o$ & on & RL or CSR & independence of cell counts\\
\hline
$\X^2_P$ and $Z_n$   & cell counts are & &   & size  $>.05$ & size  $\approx .05$ \\
under Poisson & independent Poisson  & $\widetilde \pi_{ij}=\widetilde \nu_i\,\widetilde \kappa_j$ for $i,j=1,2$ & $n$  & (i.e., liberal)& (i.e., appropriate) \\
sampling framework & variates &  & & &\\
\hline
$\X^2_P$ and $Z_n$ under & cell counts in & $N_{ii} \sim \BIN(n_i,\widetilde \pi_{ii})$  & & size  $>.05$ & size  $\approx .05$\\
row-wise binomial & each row have the same  & for $i=1,2$ with  & $n_i$ and $n$ & (i.e., liberal)& (i.e., appropriate)\\
sampling framework & binomial distribution  & $(\widetilde \pi_{11},\widetilde \pi_{12})=(\widetilde \pi_{21},\widetilde \pi_{22})$   & & &\\
\hline
$\X^2_P$ and $\widetilde{Z}_n$ under & cell counts & $(N_{11},N_{12},N_{21},N_{22}) \sim$ &  & size  $>.05$ & size  $\approx .05$ \\
overall multinomial &  have a multi-& $ \mathscr M(n,\widetilde \pi_{11},\widetilde \pi_{12},\widetilde \pi_{21},\widetilde \pi_{22})$   & $n$  & (i.e., liberal) & (i.e., appropriate) \\
sampling framework & nomial distribution & with $(\widetilde \pi_{11},\widetilde \pi_{12})=(\widetilde \pi_{21},\widetilde \pi_{22})$   & & &\\
\hline
\hline
$\X^2_D$ and $Z^D_{ij}$ & randomness in the & $\E[N_{ij}]=\frac{n_i(n_i-1)}{(n-1)} \I(i=j)$ & $n_i$ and $n$ & size  $\approx .05$ & NA \\
under RL & NN structure & $+\frac{n_i\,n_j}{(n-1)} \I(i\not=j)$    & & (i.e., appropriate) &\\
\hline
$\X^2_D$ and $Z^D_{ij}$ under & randomness in the   & $\E[N_{ij}]=\frac{n_i(n_i-1)}{(n-1)} \I(i=j)$  & $n_i$, $n$,  & size  $\approx .05$ & NA\\
CSR independence & NN structure & $+\frac{n_i\,n_j}{(n-1)} \I(i\not=j)$     &  $Q$, and $R$ & (i.e., appropriate) &\\
\hline
\end{tabular}
\caption{ \label{tab:summary-test-stat}
The summary of underlying assumptions, appropriate null hypotheses $H_o$,
and the type of conditioning for the tests based on NNCTs.
$\X^2_P$ stands for Pielou's overall test, $Z_n$ for one-sided version of Pielou's test,
$\X^2_D$ for Dixon's overall test, and $Z^D_{ij}$ for Dixon's cell-specific test.
Pielou's overall test with Yates' correction has the same properties
as the uncorrected one (hence not presented).
NA stands for ``not applicable".
}
\end{sidewaystable}


\end{document}